\documentclass[twocolumn,english,aps,pra,showpacs,floats,floatfix,footnoterule]{revtex4}
\usepackage{amsmath}
\usepackage{amsfonts}
\usepackage[T1]{fontenc}
\usepackage[latin1]{inputenc}
\usepackage{graphicx}
\usepackage{amssymb}
\usepackage{graphicx}
\usepackage{babel}
\usepackage{color}



\begin{document}

\title{Role of spatial inhomogeneity in dissociation of trapped molecular
condensates}
\author{Magnus Ögren and K. V. Kheruntsyan}
\affiliation{ARC Centre of Excellence for Quantum-Atom Optics, School of Mathematics and
Physics, University of Queensland, Brisbane, Queensland 4072, Australia}
\date{\today{}}

\begin{abstract}
We theoretically analyze dissociation of a harmonically trapped
Bose-Einstein condensate of molecular dimers and examine how the spatial
inhomogeneity of the molecular condensate affects the conversion dynamics
and the atom-atom pair correlations in the short-time limit. Both fermionic
and bosonic statistics of the constituent atoms are considered. Using the
undepleted molecular-field approximation, we obtain explicit analytic
results for the asymptotic behavior of the second-order correlation
functions and for the relative number squeezing between the dissociated
atoms in one, two and three spatial dimensions. Comparison with the
numerical results shows that the analytic approach employed here captures
the main underlying physics and provides useful insights into the dynamics
of dissociation for conversion efficiencies up to $10\%$. The results show
explicitly how the strength of atom-atom correlations and relative number
squeezing degrade with the reduction of the size of the molecular condensate.
\end{abstract}

\maketitle

\section{Introduction}

\label{sec:Introduction}

Pair production of photons has been the key mechanism in a series of
landmark experiments in quantum optics. Of particular importance have been
the experiments with pair correlated photons from an atomic radiative
cascade \cite{Aspect1981} and from parametric down-conversion \cite%
{Burnham-Weinberg1970,Ou-Mandel1987}, leading to demonstrations of
violations of Bell's inequalities and the Einstein-Podolsky-Rosen (EPR)
entanglement \cite{Aspect1982,Perrie-1985,Ou-Mandel1988,Shih1988}. Performed
initially with \emph{discrete} polarization states of photons---as a
realization of Bohm's version \cite{Bohm} of the EPR gedankenexperiment \cite%
{EPR}---the EPR correlations have been later demonstrated in their original
version, that is, for a system of observables with a \emph{continuous} spectrum
\cite{Kimble1992}. In this case, the role of a pair of canonically conjugate
variables is taken by quadrature amplitudes of spatially separated signal
and idler beams generated using a nondegenerate parametric down-conversion
in a cavity.

In atom optics, one of the simplest mechanisms for atom pair production can
be realized via dissociation of diatomic molecules \cite%
{Ketterle-diss,Rempe-diss,Greiner2005,Wieman2005,Grimm2008}. When performed
using a molecular Bose-Einstein condensate (BEC) as the initial state and
assuming that the constituent atoms are bosons, the process lends itself
\cite{Moelmer,twinbeams,Vardi-Moore,Yurovsky} as a direct matter-wave analog
of optical parametric-down conversion of an intense laser light in a $\chi
^{(2)}$-nonlinear crystal. Owing to this analogy, one can envisage that
future experiments on dissociation of molecular BECs can lead to atom optics
demonstrations of EPR correlations \cite{Kurizki,KK-EPR,Zhao-Astrakharchik}
and related tests of Bell's inequalities \cite{Fry,Hornberger}.

The most important ingredients for such experiments are: (i) the creation of
quantum degenerate samples of stable, long-lived molecules, preferably in a
rovibrational ground state (such as in recently demonstrated experiments
\cite{Cs133-Li7,Grimm87Rb2,Jin40K-87Rb}), and (ii) the development of techniques
for measuring atom-atom correlations \cite%
{Yasuda-Shimizu,Bloch-2005-2006,Greiner2005,Raizen2005,Esslinger2005,Aspect-Westbrook-2005-2007,Esteve2006,Perrin2007}%
. Demonstrations of continuous-variable EPR correlations in atomic
quadratures will additionally require the measurement of matter-wave
quadrature amplitudes using stable, mode-matched local oscillator fields as
phase references \cite{KK-EPR,Ferris2008}. This is a challenging task yet to
be realized experimentally. In the case of fermionic statistics of the
constituent atoms, dissociation of a molecular BEC offers a new paradigm of
fermionic quantum atom optics \cite{Fermidiss,Jack-Pu} and new opportunities
for extensions of fundamental tests of quantum mechanics to ensembles of
neutral fermionic atoms.

The purpose of this article is a qualitative and quantitative understanding of
the simplest density-density and atom number correlations in spontaneous
dissociation of a molecular BEC in the short-time limit. Strong atom-atom
correlations and relative number squeezing can be regarded as precursors of
more complex EPR correlations. In the short-time limit, the converted
fraction of molecules into constituent atoms is small and one can employ the
undepleted molecular-field approximation, which was successfully used in
theoretical descriptions of parametric down-conversion in quantum optics
\cite{Walls-Milburn}. Even though a less restrictive description of
dissociation dynamics can be accomplished via state-of-the-art numerical
techniques, such as the first-principles simulations using the positive-$P$
and Gaussian representations \cite%
{Savage,SavageKheruntsyanSpatial,Corney-fermionic,Midgley,Magnus-Karen-Joel-First-Principles}%
, the truncated Wigner function approach \cite{Midgley}, the
Hartree-Fock-Bogoliubov method \cite{Jack-Pu,Midgley,PMFT}, and a
generalized Gross-Pitaevskii (GP) equation \cite{Stoof-diss}, the main
motivation of the present work is to obtain approximate analytic results
which are less computationally expensive and have the intrinsic appeal of
analytic simplicity. By comparing these results with the numerical ones, one
can verify their accuracy and the range of validity in different parameter
regimes. For a related problem of atomic four-wave mixing via condensate
collisions \cite{Perrin2007}, a similar analytic treatment has been employed
in Ref. \cite{Magnus-Karen-4WM}, with the results in the short-time limit
comparing very well with the first-principles simulations in the positive-$P$ representation \cite{Perrin-theory}.

The main question that we address here is the influence of the spatial
inhomogeneity of the source (molecular) condensate on the strength of
atom-atom correlations. This is an important problem for modeling realistic
experiments performed using trapped, inhomogeneous condensates with
interactions. For example, for a harmonically trapped system, the spatial
inhomogeneity and the multimode character of the problem enters through the
shape of the density profile of the initial condensate which---in the
Thomas-Fermi (TF) limit---is close to an inverted parabola. The shape of
the source in turn determines the extent to which the effect of mode-mixing
affects the spatial structure of atom-atom correlations. For strong
inhomogeneity, the effect can be quite detrimental and can reduce the
strength of correlations significantly. Our analytic results allow us to
quantify these effects in a relatively simple way, as the short-time limits
to correlation functions are obtained in a closed explicit form in terms of
Bessel functions. In a recent rapid communication \cite%
{Magnus-Karen-dissociation-PRA-Rapid}, we presented some of these results
for dissociation in a one-dimensional (1D) geometry; in the present work, we
give the details of derivations and extend the results to 2D and 3D systems.

In addition, our comparison between the results for bosonic and fermionic
atoms allows for a demonstration of striking differences in the dissociation
dynamics that depend inherently on the difference in quantum statistics. For
short durations of dissociation, which produce low-density atomic clouds,
these differences are not obstructed by the $s$-wave scattering
interactions and have been recently studied in the context of
directionality effects due to Bose enhancement and Pauli blocking in the
dissociation of elongated molecular condensates \cite{Ogren-directionality}.

Related recent studies of molecular dissociation are concerned with the role
of confinement on the stability of the molecular BEC \cite{Vardi}, the
effect of magnetic-field fluctuations and modulations on the dissociation
dynamics near a Feshbach resonance \cite{Plata}, the dynamics of
dissociation in optical lattices \cite{Meystre-diss}, dissociation of
molecules prepared in a vortex state \cite{Molmer-twist}, connection of
dissociation with the generators of the SU(1,1) and SU(2) Lie algebras \cite%
{Vardi-SU}, loss of atom-molecule coherence due to phase diffusion \cite%
{Vardi-phase-diffusion}, as well as the use of molecular dissociation as a
probe of two-body interactions \cite{Julienne-dissociation}, collisional
resonances \cite{Rempe-Kokkelmans-dissociation}, and spectroscopic
properties of Feshbach resonance molecules \cite{Braaten-2006,Hanna-2006}.

The article is organized as follows. In Sec. \ref{sec:Model-dist} we introduce
the model for dissociation of molecular dimers consisting of either two
distinguishable bosonic atoms or two fermionic atoms in different spin
states. In Sec. \ref{sec:Model-indist} we formulate the same problem for the
case of two indistinguishable bosonic atoms. In Sec. \ref{sec:Numerics}
we present the results of a numerical analysis of molecular dissociation in
1D within the undepleted molecular-field approximation. In Sec. \ref%
{sec:Analytic} we develop an analytic approach for the short-time asymptotic
behavior and obtain explicit results for the atom-atom pair correlation
functions and the relative number squeezing in 1D, 2D, and 3D geometries,
for TF and Gaussian density profiles of the molecular BEC. Throughout these
sections, we discuss the validity of the approximate model (with no
molecular depletion) by comparing the relevant results with those obtained
using first-principles positive-$P$ simulations for bosons, in which the
molecular-field dynamics and its depletion is treated quantum mechanically.
The details of the positive-$P$ method are given in Sec. \ref%
{sec:The-effects-of-interaction}. Finally, in the same Sec. \ref%
{sec:The-effects-of-interaction} we incorporate the effects of $s$-wave
scattering interactions and analyze the system using the truncated Wigner
approach. We conclude the article with the summary Sec. \ref{sec:Summary}.

\section{Dissociation into distinguishable bosonic or fermionic atom pairs}

\label{sec:Model-dist}

To \ model the dissociations of a Bose-Einstein condensate of diatomic
molecules into pairs of constituent atoms, we start with the following
effective quantum field theory Hamiltonian, in a rotating frame \cite%
{JOptB1999-PRA2000}:%
\begin{align}
\widehat{H}& =\int d^{D}\mathbf{x}\left\{ \sum\limits_{i=0,1,2}\frac{\hbar
^{2}}{2m_{i}}|\nabla \widehat{\Psi }_{i}|^{2}+\hbar \Delta (\widehat{\Psi }%
_{1}^{\dagger }\widehat{\Psi }_{1}+\widehat{\Psi }_{2}^{\dagger }\widehat{%
\Psi }_{2})\right.  \notag \\
& \left. -i\hbar \chi \left( \widehat{\Psi }_{0}^{\dagger }\widehat{\Psi }%
_{1}\widehat{\Psi }_{2}-\widehat{\Psi }_{2}^{\dagger }\widehat{\Psi }%
_{1}^{\dagger }\widehat{\Psi }_{0}\right) \right\} .  \label{hamiltonian}
\end{align}%
Here we assume that the molecules [described by the field operator $\widehat{%
\Psi }_{0}(\mathbf{x},t)$] are made of either two distinguishable bosonic
atoms or two fermionic atoms in different spin states. In both cases, $%
\widehat{\Psi }_{0}(\mathbf{x},t)$ is a bosonic field operator satisfying
the standard commutation relation $[\widehat{\Psi }_{0}(\mathbf{x},t),%
\widehat{\Psi }_{0}^{\dagger }(\mathbf{x}^{\prime },t)]=\delta ^{D}(\mathbf{x%
}-\mathbf{x}^{\prime })$, with $D=1,2$ or $3$ corresponding to the
dimensionality of the system. The atomic field operators, $\widehat{\Psi }%
_{i}(\mathbf{x},t)$ ($i=1,2$), satisfy either bosonic commutation or
fermionic anticommutation relations, $[\widehat{\Psi }_{i}(\mathbf{x},t),%
\widehat{\Psi }_{j}^{\dagger }(\mathbf{x}^{\prime },t)]=\delta _{ij}\delta
^{D}(\mathbf{x}-\mathbf{x}^{\prime })$ or $\{\widehat{\Psi }_{i}(\mathbf{x}%
,t),\widehat{\Psi }_{j}^{\dagger }(\mathbf{x}^{\prime },t)\}=\delta
_{ij}\delta ^{D}(\mathbf{x}-\mathbf{x}^{\prime })$, depending on the
underlying statistics.

The first term in the Hamiltonian (\ref{hamiltonian}) describes the kinetic
energy where the atomic masses are $m_{1}$ and $m_{2}$, whereas the
molecular mass is $m_{0}=m_{1}+m_{2}$. For simplicity, we consider the
case of equal atomic masses (same isotope atoms), with $m_{1}=m_{2}\equiv m$
and $m_{0}=2m$.

The coupling constant $\chi \equiv \chi _{D}$ is responsible for coherent
conversion of molecules into atom pairs, for example, via optical Raman transitions,
an rf transition, or a Feshbach resonance sweep (see, for example, Refs.
\cite%
{JOptB1999-PRA2000,PDKKHH-1998,Superchemistry,Feshbach-KKPD,Timmermans,JJ-1999,Holland}
and \cite{Stoof-review,Julienne-review} for recent reviews); the microscopic
expressions for $\chi $ in 1D, 2D, and 3D can be found in Ref. \cite{PMFT}.

The detuning $\Delta $ is defined to give the overall energy mismatch $%
2\hbar \Delta $ between the free two-atom state in the dissociation
threshold and the bound molecular state (including the relative frequencies
of the Raman lasers or the frequency of the rf field; for further details,
see Refs. \cite{twinbeams,PMFT}). Unstable molecules, spontaneously
dissociating into pairs of constituent atoms, correspond to $\Delta <0$,
with $2\hbar |\Delta |$ being the total dissociation energy.

The trapping potential for preparing the initial molecular BEC---with any
residual atoms being removed---is omitted from the Hamiltonian since we
assume that once the dissociation is invoked, the trapping potential is
switched off, so that the dynamics of dissociation is taking place in free
space (in 1D and 2D geometries, we assume that the confinement in the
eliminated dimensions is kept on so that the free-space dynamics refers only
to the relevant dimension under consideration). We assume that the switching
on of the atom-molecule coupling and switching off of the trapping potential
is done in the regime of a sudden jump \cite{Hanna-2006}. Accordingly, the
preparation stage is reduced to assuming a certain initial state of the
molecular BEC in a trap, after which the dynamics is governed by the
Hamiltonian (\ref{hamiltonian}).

In what follows we initially treat the dynamics of dissociation in the
undepleted molecular condensate approximation in which the molecules are
represented as a fixed classical field. The undepleted molecular
approximation is valid for short-enough dissociation times during which the
converted fraction of molecules does not exceed $\sim 10\%$ \cite%
{Savage,PMFT,Ogren-directionality}. In this regime the dissociation
typically produces low-density atomic clouds for which the atom-atom $s$%
-wave scattering interactions are negligible \cite{Savage}; hence, their
absence from our Hamiltonian. Additionally, the atom-molecule interactions
will initially appear as an effective spatially dependent detuning due to
the mean-field interaction energy; this can be neglected by operating at
relatively large absolute values of the dissociation detuning $|\Delta |$ so
that the total dissociation energy $2\hbar |\Delta |$ dominates the
mean-field energy shifts \cite{Savage,Ogren-directionality}. As means of
verifying the regime of validity of these approximations, in Sec. \ref%
{sec:The-effects-of-interaction} we incorporate the effects of molecular
depletion and $s$-wave scattering interactions using the positive-$P$
representation and the truncated Wigner method. As these numerical methods
are only applicable to bosons, our comparison and conclusions are restricted
to the case of dissociation into bosonic atoms. For the case of fermionic
atoms, the development of stochastic methods that may facilitate similar
comparison in the future are under development \cite%
{Magnus-Karen-Joel-First-Principles}.

\subsection{Heisenberg equations in the undepleted molecular condensate
approximation}

The undepleted molecular-field approximation is invoked as follows. Assuming
that the molecules are in a coherent state initially, with the density
profile $\rho _{0}(\mathbf{x})$ given by the ground-state solution of the
standard GP equation in a harmonic trap, we replace the molecular-field
operator by its coherent mean-field amplitude, $\widehat{\Psi }_{0}(\mathbf{x%
},0)\rightarrow \langle \widehat{\Psi }_{0}(\mathbf{x},0)\rangle =\Psi _{0}(%
\mathbf{x},0)=\sqrt{\rho _{0}(\mathbf{x})}$, which we assume is real without
loss of generality. We can next introduce an effective, spatially dependent
coupling,
\begin{equation}
g(\mathbf{x})=\chi \sqrt{\rho _{0}(\mathbf{x})},
\end{equation}%
and write the Heisenberg equations for the atomic fields as follows:
\begin{align}
\frac{\partial \widehat{\Psi }_{1}(\mathbf{x},t)}{\partial t}& =i\left[
\frac{\hbar }{2m}\nabla ^{2}-\Delta \right] \widehat{\Psi }_{1}(\mathbf{x}%
,t)\pm g(\mathbf{x})\widehat{\Psi }_{2}^{\dag }(\mathbf{x},t),  \notag \\
&  \label{Heisenberg-eqs} \\
\frac{\partial \widehat{\Psi }_{2}^{\dagger }(\mathbf{x},t)}{\partial t}& =-i%
\left[ \frac{\hbar }{2m}\nabla ^{2}-\Delta \right] \widehat{\Psi }%
_{2}^{\dagger }(\mathbf{x},t)+g(\mathbf{x})\widehat{\Psi }_{1}(\mathbf{x},t).
\notag
\end{align}%
Here and hereafter the $+$ and $-$ signs (in general, upper and lower signs)
are for bosonic and fermionic atoms, respectively.

Transforming to Fourier space, $\widehat{\Psi }_{j}(\mathbf{x},t)=\int d^{D}%
\mathbf{k}\widehat{a}_{j}(\mathbf{k},t)\exp (i\mathbf{k\cdot x})/(2\pi
)^{D/2}$, where the amplitude operators $\widehat{a}_{j}(\mathbf{k},t)$
satisfy commutation or anticommutation relations, $[\widehat{a}_{i}(\mathbf{k%
}),\widehat{a}_{j}^{\dagger }(\mathbf{k}^{\prime })]=\delta _{ij}\delta ^{D}(%
\mathbf{k}-\mathbf{k}^{\prime })$ or $\{\widehat{a}_{i}(\mathbf{k}),\widehat{%
a}_{j}^{\dagger }(\mathbf{k}^{\prime })\}=\delta _{ij}\delta ^{D}(\mathbf{k}-%
\mathbf{k}^{\prime })$, according to the underlying statistics, we can
rewrite Eqs.~(\ref{Heisenberg-eqs}) as a set of linear operator equations:%
\begin{eqnarray}
\frac{d\widehat{a}_{1}(\mathbf{k},t)}{dt} &=&-i\Delta _{k}\widehat{a}_{1}(%
\mathbf{k},t)\pm \int \frac{d^{D}\mathbf{q}}{(2\pi )^{D/2}}\widetilde{g}(%
\mathbf{q}+\mathbf{k})\widehat{a}_{2}^{\dagger }(\mathbf{q},t),  \notag \\
&&  \label{HeisenbergsEquation} \\
\frac{d\widehat{a}_{2}^{\dagger }(\mathbf{k},t)}{dt} &=&i\Delta _{k}\widehat{%
a}_{2}^{\dagger }(\mathbf{k},t)+\int \frac{d^{D}\mathbf{q}}{(2\pi )^{D/2}}%
\widetilde{g}(\mathbf{q}-\mathbf{k})\widehat{a}_{1}(-\mathbf{q},t).  \notag
\end{eqnarray}%
Here%
\begin{equation}
\widetilde{g}(\mathbf{k})=\frac{1}{(2\pi )^{D/2}}\int d^{D}\mathbf{x}e^{-i%
\mathbf{k\cdot x}}g(\mathbf{x}),  \label{FourierCoefficients}
\end{equation}%
is the Fourier transform of the effective coupling $g(\mathbf{x})$, and we
have defined $\Delta _{k}\equiv \hbar k^{2}/(2m) +\Delta $, where $k=|%
\mathbf{k}|$.

The general structure of solutions following from Eqs. (\ref%
{HeisenbergsEquation}), with vacuum initial conditions for the atomic
fields, can be easily understood if we rewrite these operator equations in
terms of ordinary differential equations for all possible second-order
moments of the atomic field operators. By doing so, one can show that the
equations for the normal and anomalous densities, $\langle \widehat{a}%
_{j}^{\dagger }(\mathbf{k},t)\widehat{a}_{j}(\mathbf{k}^{\prime },t)\rangle $
and $\langle \widehat{a}_{1}(\mathbf{k},t)\widehat{a}_{2}(\mathbf{k}^{\prime
},t)\rangle $, together with their complex conjugates, form a closed set and
develop nonzero populations from the delta-function \textquotedblleft
seed\textquotedblright\ terms that originate from the following identity: $%
\langle \widehat{a}_{j}(\mathbf{k},t)\widehat{a}_{j}^{\dagger }(\mathbf{k}%
^{\prime },t)\rangle =\delta ^{D}(\mathbf{k}-\mathbf{k}^{\prime })\pm
\langle \widehat{a}_{j}^{\dagger }(\mathbf{k},t)\widehat{a}_{j}(\mathbf{k}%
^{\prime },t)\rangle $. The other second-order moments, $\langle \widehat{a}%
_{1}^{\dagger }(\mathbf{k},t)\widehat{a}_{2}(\mathbf{k}^{\prime },t)\rangle $
and $\langle \widehat{a}_{j}(\mathbf{k},t)\widehat{a}_{j}(\mathbf{k}^{\prime
},t)\rangle $, also form a closed set; however, they never develop nonzero
populations if the populations were absent initially. According to this
structure, the only nonzero second-order moments are the normal and
anomalous densities%
\begin{gather}
n_{j}(\mathbf{k},\mathbf{k}^{\prime },t)\equiv \langle \widehat{a}%
_{j}^{\dagger }(\mathbf{k},t)\widehat{a}_{j}(\mathbf{k}^{\prime },t)\rangle
,\;\;j=1,2, \\
m_{12}(\mathbf{k},\mathbf{k}^{\prime },t)\equiv \langle \widehat{a}_{1}(%
\mathbf{k},t)\widehat{a}_{2}(\mathbf{k}^{\prime },t)\rangle ,
\end{gather}%
with $n_{1}(\mathbf{k},\mathbf{k}^{\prime },t)\!=\!n_{2}(\mathbf{k},\mathbf{k%
}^{\prime },t)$, whereas $\langle \widehat{a}_{1}^{\dagger }(\mathbf{k},t)%
\widehat{a}_{2}(\mathbf{k}^{\prime },t)\rangle \!=\!0$ and $\langle \widehat{%
a}_{j}(\mathbf{k},t)\widehat{a}_{j}(\mathbf{k}^{\prime },t)\rangle \!=\!0$.
Since the effective Hamiltonian corresponding to Eqs. (\ref{Heisenberg-eqs})
is quadratic in the field operators, any higher-order moments or expectation
values of products of creation and annihilation operators will factorize
according to Wick's theorem into products of the normal and anomalous
densities $n_{j}(\mathbf{k},\mathbf{k}^{\prime },t)$ and $m_{12}(\mathbf{k},%
\mathbf{k}^{\prime },t)$.

\section{Dissociation into indistinguishable bosonic atoms}

\label{sec:Model-indist}

For completeness, we also analyze dissociation of a BEC of molecular
dimers made of pairs of indistinguishable bosonic atoms in the same spin
state. This is described by the following effective Hamiltonian \cite%
{PDKKHH-1998}, in a rotating frame:
\begin{align}
\widehat{H}& =\int d^{D}\mathbf{x}\left\{ \sum\limits_{i=0,1}\frac{\hbar ^{2}%
}{2m_{i}}|\nabla \widehat{\Psi }_{i}|^{2}+\hbar \Delta \widehat{\Psi }%
_{1}^{\dagger }\widehat{\Psi }_{1}\right.  \notag \\
& \left. -i\frac{\hbar \chi }{2}\left( \widehat{\Psi }_{0}^{\dagger }%
\widehat{\Psi }_{1}^{2}-\widehat{\Psi }_{1}^{\dagger 2}\widehat{\Psi }%
_{0}\right) \right\} .  \label{Hindistinguish}
\end{align}%
Here $\widehat{\Psi }_{1}(\mathbf{x,}t)$ is the atomic field operator, $\chi
$ is the respective atom-molecule coupling \cite{PMFT,Feshbach-KKPD}, $m_{0}$
and $m_{1}\equiv m$ are, respectively, the molecular and atomic masses ($m_{0}=2m$), and $%
\Delta $ is the detuning corresponding to the total dissociation energy of $%
2\hbar |\Delta |$.

The treatment of this system is essentially the same as in the previous case
of distinguishable atoms, except that the field operators $\widehat{\Psi }%
_{2}^{\dag }(\mathbf{x},t)$ and $\widehat{a}_{2}^{\dagger }(\mathbf{k},t)$
in Eqs. (\ref{Heisenberg-eqs}) and (\ref{HeisenbergsEquation}) are replaced,
respectively, with $\widehat{\Psi }_{1}^{\dag }(\mathbf{x},t)$ and $\widehat{a}%
_{1}^{\dagger }(\mathbf{k},t)$. The corresponding Heisenberg equations now
read as
\begin{equation}
\frac{\partial \widehat{\Psi }_{1}(\mathbf{x},t)}{\partial t}=i\left[ \frac{%
\hbar \nabla ^{2}}{2m}-\Delta \right] \widehat{\Psi }_{1}(\mathbf{x},t)+g(%
\mathbf{x})\widehat{\Psi }_{1}^{\dag }(\mathbf{x},t),
\label{Heisnberg-indistinguishable}
\end{equation}%
and
\begin{equation}
\frac{d\widehat{a}_{1}(\mathbf{k},t)}{dt}=-i\Delta _{k}\widehat{a}_{1}(%
\mathbf{k},t)+\int \frac{d^{D}\mathbf{q}}{(2\pi )^{D/2}}\widetilde{g}(%
\mathbf{q}+\mathbf{k})\widehat{a}_{1}^{\dagger }(\mathbf{q},t),
\label{Heisenberg-indistinguish-F}
\end{equation}%
whereas the nonzero normal and anomalous densities are given by
\begin{gather}
n_{1}(\mathbf{k},\mathbf{k}^{\prime },t) \equiv \langle \widehat{a}%
_{1}^{\dagger }(\mathbf{k},t)\widehat{a}_{1}(\mathbf{k}^{\prime },t)\rangle ,
\\
m_{11}(\mathbf{k},\mathbf{k}^{\prime },t) \equiv \langle \widehat{a}_{1}(%
\mathbf{k},t)\widehat{a}_{1}(\mathbf{k}^{\prime },t)\rangle ,
\end{gather}%
where we have omitted the atomic spin index for notational simplicity.

\section{Numerical results and discussion}

\label{sec:Numerics}

In our numerical treatment of the problem (present section) we only
consider a 1D system. This is to make the problem computationally tractable.
However, based on the physical understanding that we develop, we expect our
results to be at least qualitatively valid for 3D systems, as is the case
for a 2D problem treated recently in Ref. \cite{Ogren-directionality}. The
analytical results of Sec. \ref{sec:Analytic}, on the other hand, are
obtained for 1D, 2D and 3D systems.

From the structure of Eqs. (\ref{HeisenbergsEquation}) we can easily
recognize the role of mode-mixing in the spatially inhomogeneous treatment
compared to the case of a uniform molecular condensate. In the uniform case,
the Fourier transform of the effective coupling $g_{0}$ is a delta function $%
\widetilde{g}(\mathbf{k})=(2\pi )^{D/2}g_{0}\delta (\mathbf{k})$, so that
the operator $\widehat{a}_{1}(\mathbf{k})$ couples to the conjugate of the
partner spin component at exactly the opposite momentum, $\widehat{a}%
_{2}^{\dagger }(-\mathbf{k})$. Therefore, the entire set of coupled
equations breaks down into pairs of equations that couple only the opposite
momentum components of the two atomic fields with different spins.
Accordingly, only the diagonal and antidiagonal terms of the normal and
anomalous densities, $n_{j}( \mathbf{k},\mathbf{k})$ and $m_{12}(\mathbf{k},-%
\mathbf{k})$, develop nonzero populations as the dissociation proceeds.

In the present inhomogeneous case, on the other hand, the finite width of
the effective coupling $\widetilde{g}(\mathbf{k})$ implies that $\widehat{a}%
_{1}(\mathbf{k})$ couples not only to $\widehat{a}_{2}^{\dagger }(-\mathbf{k}%
)$ in Eq.~(\ref{HeisenbergsEquation}), but also to a range of momenta in the
neighborhood of $-\mathbf{k}$, within $-\mathbf{k}\pm \delta \mathbf{k}$.
This is the origin of mode-mixing. The spread in $\delta \mathbf{k}$
determines the width of the pair correlation between the atoms in the two
opposite spin states that have equal but opposite momenta. The width is
ultimately related to the momentum width of the source molecular condensate,
as we show later in this article.

At a qualitative level, the finite width of the pair correlation at opposite
momenta can be understood from a simple momentum conservation argument. For
a molecule at rest, the dissociation produces one atom in each spin state
satisfying $\mathbf{k}_{1}+\mathbf{k}_{2}=0$, and therefore $\mathbf{k}_{2}=-%
\mathbf{k}_{1}$. From energy conservation $2\hbar |\Delta |=(\hbar ^{2}|%
\mathbf{k}_{1}|^{2}+\hbar ^{2}|\mathbf{k}_{2}|^{2})/2m$, the absolute
momentum of each atom is given by $k_{0}=|\mathbf{k}_{1}|=|\mathbf{k}_{2}|=%
\sqrt{2m|\Delta |/\hbar }$. The same momentum conservation holds in the
center-of-mass frame of the molecule if it has a finite momentum offset $%
\delta \mathbf{k}$ due to the initial momentum spread of the condensate. In
the laboratory frame, this center-of-mass momentum offset leads to an offset
from $\pm \mathbf{k}_{0}$ in the momenta of dissociated atoms, $\mathbf{k}%
_{1}=\mathbf{k}_{0}+\delta \mathbf{k}/2$ and $\mathbf{k}_{2}=-\mathbf{k}%
_{0}+\delta \mathbf{k}/2$. This implies that a pair of atom detectors set to
detect atoms with these momenta will produce a positive pair correlation
signal and will therefore contribute to the finite width of the order of $%
\delta \mathbf{k}$ in the density-density correlation function \cite%
{Comment2}.

We now turn to the quantitative analysis of atom-atom correlations. Before
we proceed, however, we discuss the role of mode-mixing on simpler
observables---the atomic momentum distribution and mode population dynamics
in the two spin states.

\subsection{Momentum distribution and mode population dynamics}

In a finite quantization volume, the wave vector $\mathbf{k}$ is discrete
and the plane-wave mode annihilation and creation operators $\widehat{a}_{j,%
\mathbf{k}}=\widehat{a}_{j}(\mathbf{k},t)(\Delta k_{x}\Delta k_{y}\Delta
k_{z})^{1/2}$ and $\widehat{a}_{j,\mathbf{k}}^{\dagger }=\widehat{a}%
_{j}^{\dagger }(\mathbf{k},t)(\Delta k_{x}\Delta k_{y}\Delta k_{z})^{1/2}$
(where $j=1,2$ and $\Delta k_{x,y,z}$ are the lattice spacings in $x$, $y$,
and $z$ directions) may be organized into a vector $\vec{\hat{a}}$. The
Heisenberg equations (\ref{HeisenbergsEquation}) may then be written in
vector-matrix form as $d\vec{\hat{a}}/dt=\mathbf{M}\vec{\hat{a}}$, where $%
\mathbf{M}$ is a square matrix of complex numbers of dimension equal to
twice the total number of lattice points. The solutions of these linear
operator equations can be found by numerically computing the matrix
exponential $\exp (\mathbf{M}t)$. The task is relatively simple in 1D, which
is the case that we present here.

\begin{figure}[tbp]
\includegraphics[height=5cm]{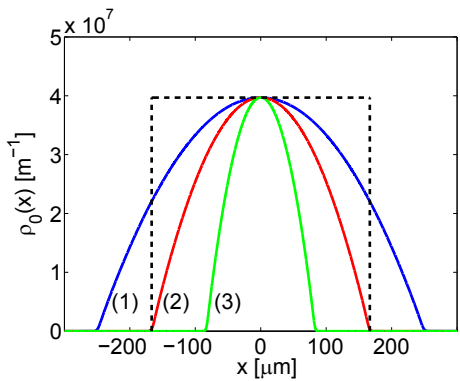} 
\caption{(Color online) Molecular BEC density profiles $\protect\rho_{0}(x)$ as ground-state solutions of the Gross-Pitaevskii equation in a harmonic trap with
longitudinal frequencies $\protect\omega /2\protect\pi =1$, $3/2$, and $3$
Hz represented, respectively, by the curves (1), (2), and (3). In the
Thomas-Fermi (TF) regime, the corresponding TF radii are (1) $R_{TF}\simeq
250$ $\protect\mu $m, (2) $R_{TF}\simeq 167$ $\protect\mu $m, and (3) $%
R_{TF}\simeq 83$ $\protect\mu $m. The dashed box illustrates a uniform
system which is size matched with the inhomogeneous system (1). Other
physical parameters are given in Ref.~\protect\cite{PhysicalParameters}. }
\label{figDifferentWidths}
\end{figure}

In our numerical analysis, we consider three typical examples of the density
profiles $\rho _{0}(x)$ of the molecular BEC, corresponding to relatively
weak, intermediate, and strong inhomogeneity. These are shown in Fig. \ref%
{figDifferentWidths} and correspond to having different frequencies of the
longitudinal trapping potential along $x$ and the same peak density $%
\rho_{0}\equiv \rho _{0}(0)$. The density profiles shown are given by the
ground-state solution of the 1D GP equation in a harmonic trap and can be
closely approximated by a TF inverted parabola $\rho
_{0}(x)=\rho _{0}(1-x^{2}/R_{TF}^{2})$ for $|x|<R_{TF}$ and $\rho_{0}(x)=0$
elsewhere; the three examples shown correspond to the TF radii of $%
R_{TF}=250$ $\mu $m, $R_{TF}=167$ $\mu $m, and $R_{TF}=83$ $\mu $m.

\begin{figure}[tbp]
~~\includegraphics[width=6.2cm]{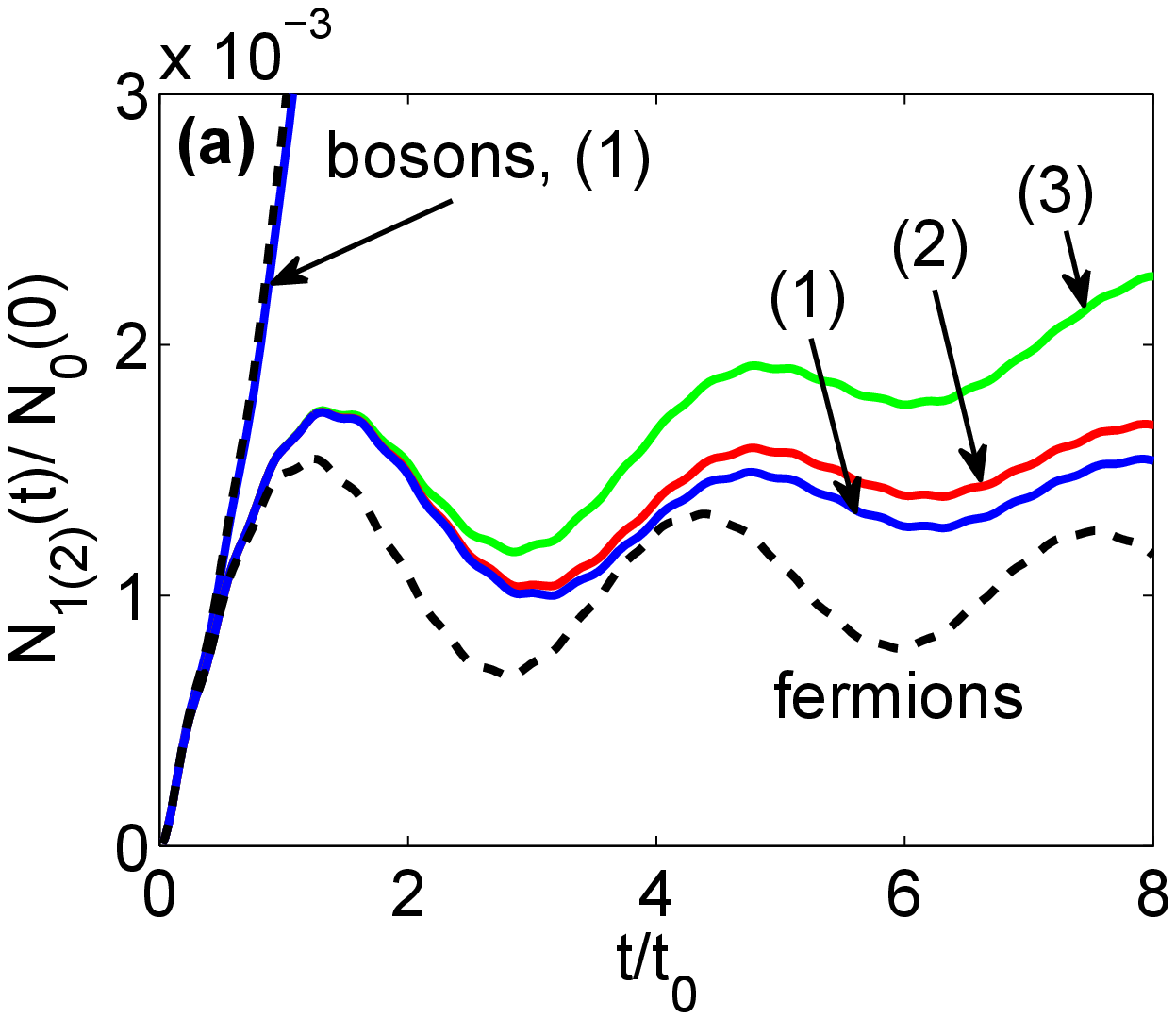}
\par
\includegraphics[width=6.4cm]{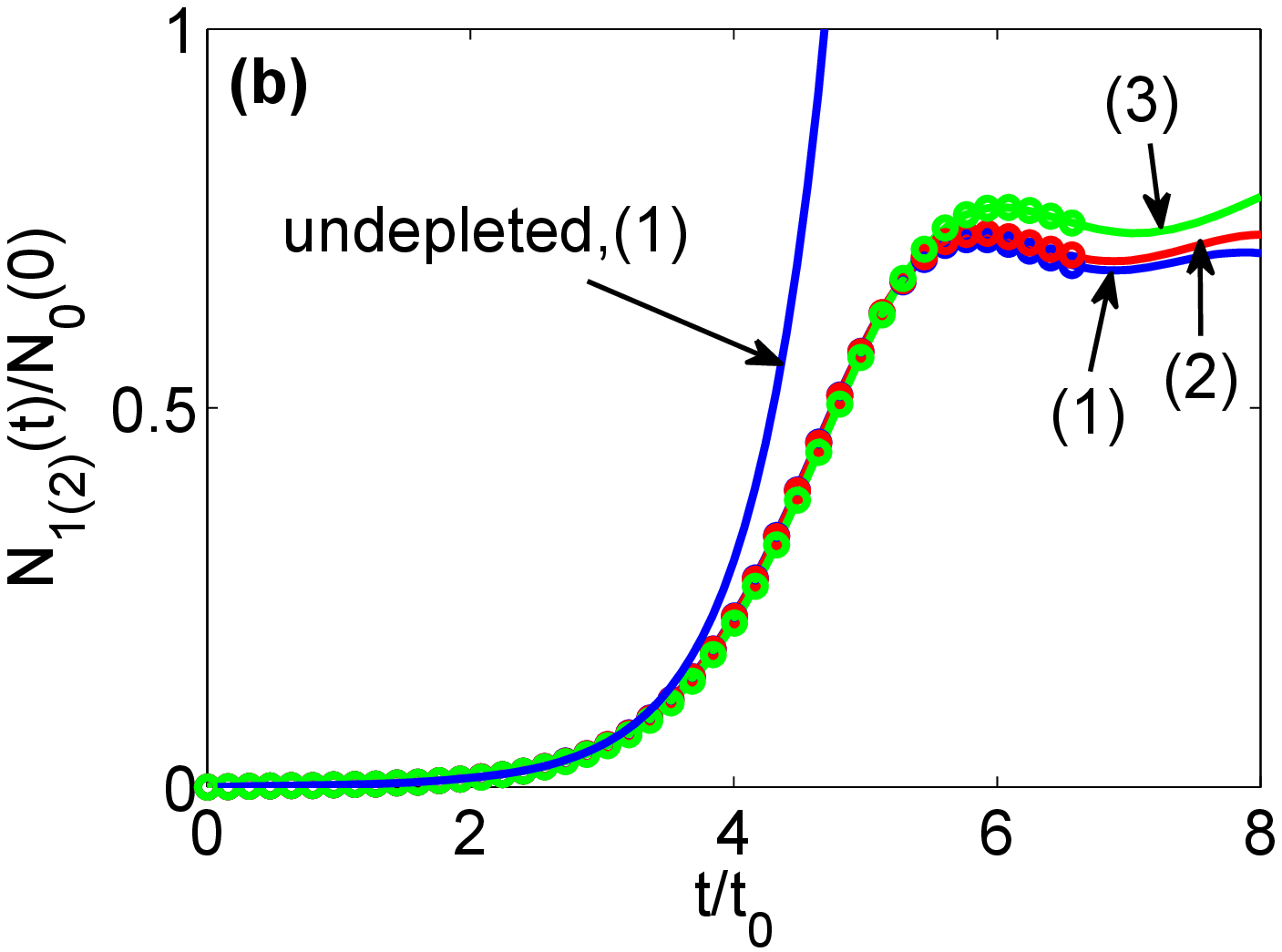}
\caption{(Color online) (a) Total fractional number of atoms $N_{j}(t)/N_{0}(0)$ [$%
N_{1}(t)=N_{2}(t)$] as a function of time $t/t_{0}$ in one of the spin
components in dissociation into fermionic or bosonic atoms. The different
curves (1), (2), and (3) correspond to dissociation of molecular BECs having
density profiles as indicated in Fig. \protect\ref{figDifferentWidths}, with
the dashed line referring to the uniform case. The results are obtained
within the undepleted molecular approximation. The dimensionless detuning $%
\protect\delta =\Delta t_{0}$ is $\protect\delta =-9$, where the time scale
is $t_{0}=1/\protect\chi \protect\sqrt{\protect\rho_{0}}$. For $\protect\chi %
=3.15\times 10^{-2}$ m$^{1/2}/$s and $\protect\rho_{0}=4\times 10^{7}$ m$%
^{-1}$, the time scale would be $t_{0}=5$ ms. Accordingly, the absolute
detuning is $|\Delta |=1800$ s$^{-1}$ and the resonant momentum is $k_{0}=%
\protect\sqrt{2m|\Delta |/\hbar }\simeq 1.5\times 10^{6}$ m$^{-1}$ in the
present example \protect\cite{PhysicalParameters}. (b) Total fractional
numbers of bosonic atoms $N_{j}(t)/N_{0}(0)$ from numerical simulations that
take into account molecular depletion. The different curves are as in (a),
with the curve (1) from the undepleted molecular approximation shown for
comparison. The solid curves are from the truncated Wigner method, while the
circles are from the exact positive-$P$ method (see Sec. \protect\ref%
{sec:The-effects-of-interaction}), which in this example is limited to
simulation duration of $t/t_{0}\sim 6.5$ due to the growing sampling errors.
}
\label{total-atom-number}
\end{figure}

To allow for a comparison of the present nonuniform treatment with the known
analytic solutions of a uniform model \cite{Fermidiss}, we also show a
uniform box system (dashed line in Fig. \ref{figDifferentWidths}), which is
size-matched with the largest trapped system, curve (1). We choose the
size-matched uniform box to have the same uniform density $\rho _{u}$ as the
peak density $\rho_{0}$ of the nonuniform system and the same total initial
number of molecules $N_{0}$. For a simple TF parabola, $N_{0}$ is given by $%
N_{0}=4\rho_{0}R_{TF}/3$, while for the box system $N_{0}=\rho _{u}L$, and
therefore we require that the box length is $L=4R_{TF}/3$.

In Fig. \ref{total-atom-number}(a) we plot the total number of atoms in each
spin component,%
\begin{equation}
N_{j}(t) =\int dkn_{j}(k,t),  \label{NumberOfAtoms}
\end{equation}%
where
\begin{equation}
n_{j}(k,t)=n_{j}(k,k,t)=\langle \widehat{a}_{j}^{\dagger }(k,t)\widehat{a}%
_{j}(k,t)\rangle
\end{equation}%
is the density distribution in momentum space. The three curves (1), (2), and
(3) referring to fermionic atoms correspond, respectively, to dissociation
of molecular condensates with the density profiles shown Fig. \ref%
{figDifferentWidths}. Plotted are the fractional atom numbers, $%
N_{j}(t)/N_{0}(0)$ [with $N_{1}(t)=N_{2}(t)$], where $N_{0}(0)$ in each case
is the respective total initial number of molecules. For comparison, we also
show the result for bosonic atoms corresponding to the molecular density
profile (1). The two dashed curves are the respective (fermionic or bosonic)
results for a uniform system, size matched with the molecular condensate
profile (1).

The differences between the curves (1), (2), and (3) demonstrate the strong
dependence of the dissociation dynamics on the inhomogeneity of the initial
molecular condensate. We note that all fermionic examples are in the
parameter regime where the dynamics is dominated by Pauli blocking of
individual atomic modes rather than by molecular depletion \cite{PMFT}.
Accordingly, only a small fraction of molecules is converted into atoms and
this justifies the use of the undepleted molecular approximation for longer
time scales than in the respective bosonic systems.

In Fig. \ref{total-atom-number}(b) we show the results of numerical
simulations that go beyond the undepleted molecular approximation. The
simulations are performed for dissociation into bosonic atoms using the
positive-$P$ representation and the truncated Wigner methods (see Sec. \ref%
{sec:The-effects-of-interaction} for further details). They take into
account the conversion dynamics and quantum fluctuations of the molecular
field, in contrast to the results of Fig. \ref{total-atom-number}(a). As we
see, the results of the undepleted molecular approximation are in excellent
agreement with the exact positive-$P$ results for dissociation durations up
to $t/t_{0}\sim 3.5$ corresponding to more than $10$\% conversion. At later
times, the exact results show the slowing down of the atom number growth due
to the depletion of the molecular condensate, followed by the reverse
process of atom-atom recombination into molecules.

\begin{figure}[tbp]
\includegraphics[height=5.5cm]{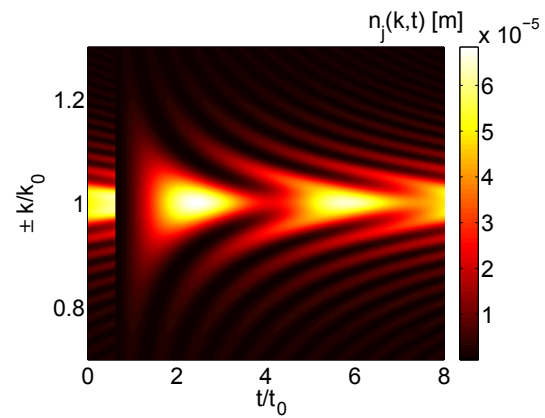} 
\caption{(Color online) Atomic momentum distribution $n_{j}(k,t)$ [$%
n_{1}(k,t)=n_{2}(k,t)$] in one of the spin components for dissociation into
fermionic atoms, as a function of a scaled time $t/t_{0}$. The example is
for the molecular BEC profile corresponding to the case (1) in Fig. \protect
\ref{figDifferentWidths}. The dimensionless detuning is $\protect\delta =-9$,
as in Fig. \protect\ref{total-atom-number}. }
\label{spectrum}
\end{figure}

To further illustrate the differences between the uniform and nonuniform
results, we plot in Fig. \ref{spectrum} the momentum distribution of\ the
dissociated fermionic atoms, $n_{j}(k,t)$, as a function of time. The
distribution is symmetric around the origin and has two peaks centered
around the resonant momenta $k=\pm k_{0}$; in Fig. \ref{spectrum} we only
show the spectrum around $k=k_{0}$. Qualitatively, the momentum distribution
is similar to the one obtained within the uniform treatment except that the
oscillation of the resonant momentum is no longer periodic and the minima do
not reach zero \cite{Fermidiss,PMFT}. We recall that in the uniform case
the oscillations of different plane-wave mode occupancies are periodic and
are given by \cite{Fermidiss}
\begin{equation}
n_{j,k}(t)=\frac{g_{0}^{2}}{g_{0}^{2}+\Delta _{k}^{2}}\sin ^{2}(\sqrt{%
g_{0}^{2}+\Delta _{k}^{2}}\ t),
\end{equation}%
where $g_{0}\equiv g(0)$ is the effective coupling and $\Delta _{k}\equiv
\hbar k^{2}/(2m)+\Delta =\hbar (k^{2}-k_{0}^{2})/2m$. In the bosonic case,
we find that the density distributions are closer to the respective uniform
results of Ref. \cite{Savage,Fermidiss}, $n_{j,k}(t)=[g_{0}^{2}/(g_{0}^{2}-%
\Delta _{k}^{2})]\sinh ^{2}(\sqrt{g_{0}^{2}-\Delta _{k}^{2}}\ t) $, and we
do not show them here.

\begin{figure}[tbp]
\includegraphics[width=8.7cm]{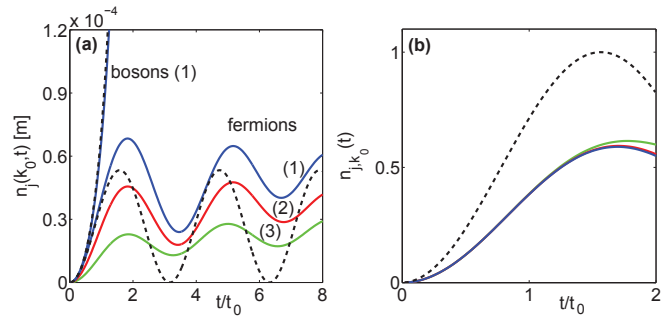}
\caption{(Color online) (a) Peak density $n_{j}(k_{0},t)$ [$n_{1}(k_{0},t)=n_{2}(k_{0},t)$]
as a function of time $t/t_{0}$ in one of the spin components in
dissociation into fermionic or bosonic atoms. The three different curves for
fermions correspond to the molecular BEC profiles (1), (2), and (3) of Fig.
\protect\ref{figDifferentWidths}, while the oscillating dashed curve is the
peak atomic density for a uniform system that is size matched with (1). The
bosonic curve together with the respective dashed curve for a size-matched
uniform system is for the molecular BEC profile (1) of Fig. \protect\ref%
{figDifferentWidths}. The dimensionless detuning is the same as in Fig.
\protect\ref{total-atom-number}, $\protect\delta =-9$. (b) Comparison of the
average mode \emph{occupancy} in the uniform fermionic system (dashed curve)
and the respective nonuniform systems (solid curves), corresponding to cases
(1), (2), and (3) of Fig. \protect\ref{figDifferentWidths}. The latter three
curves are almost indistinguishable from each other on this scale. }
\label{mode-dynamics}
\end{figure}

In Fig. \ref{mode-dynamics}(a) we monitor the atomic density at the resonant
momentum $k_{0}$ as a function of time; the three fermionic curves
correspond, respectively, to the molecular density profiles (1), (2), and (3)
of Fig. \ref{figDifferentWidths}, whereas the oscillatory dashed line
corresponds to the analytic solution for a size-matched uniform system (1).
The analytic result represents the density distribution $n_{j}(k_{0},t)$
and is obtained from the average occupation number of the resonant mode, $%
n_{j,k_{0}}(t)=\sin ^{2}(g_{0}t)$, by converting it into the density
distribution via $n_{j}(k_{0},t)=n_{j,k_{0}}(t)/\Delta k$, where $\Delta
k=2\pi /L$ is the mode spacing of the size-matched uniform system of length $%
L$. The respective bosonic results for the largest molecular BEC are also
shown for comparison; in this case the uniform analytic result for the
resonant mode is given by $n_{j,k_{0}}(t)=\sinh ^{2}(g_{0}t)$ \cite%
{Savage,Fermidiss} and grows exponentially with time due to Bose stimulation.

An alternative way of comparing the results for the uniform and nonuniform
systems is to define the physical \textquotedblleft modes\textquotedblright\
of the nonuniform system and to compare their average occupation numbers
with those obtained in the uniform finite size box. In the later case, the
natural modes of the system are the plane-wave modes which coincide with our
computational lattice modes. In the case of dissociation into fermionic
atoms, the plane-wave mode occupation numbers are bound to be no more than $%
1 $ by the Pauli exclusion principle \cite{Fermidiss}. In the nonuniform
system, on the other hand, the plane-wave modes are not the natural modes of
the system and therefore care should be taken when defining the physical
modes and discussing the Pauli exclusion principle. Following Glauber's
theory of optical coherence in the context of matter waves \cite{Glauber},
we define the atomic \textquotedblleft mode\textquotedblright\ in the
nonuniform system using the first-order coherence length, $\Delta k^{(%
\mathrm{coh})}$, which in turn is defined via the first-order correlation
function%
\begin{equation}
g_{jj}^{(1)}(k,k^{\prime },t)=\frac{\langle \widehat{a}_{j}^{\dagger }(k,t)%
\widehat{a}_{j}(k^{\prime },t)\rangle }{\sqrt{n_{j}(k,t)n_{j}(k^{\prime },t)}%
}.
\end{equation}%
More specifically, we define $\Delta k^{(\mathrm{coh})}$ as the distance $%
|k-k^{\prime }|$ over which $|g_{jj}^{(1)}(k,k^{\prime },t)|$ reduces by a
factor of $2$ from its peak value at $k-k^{\prime }=0$. The average mode
occupation is then given by $n_{j,k}(t)=n_{j}(k,t)\Delta k^{(\mathrm{coh})}$%
. In Fig. \ref{mode-dynamics}(b) we plot $n_{j,k_{0}}(t)$, defined in this
way, and compare it with the occupancies of the plane-wave modes of the
uniform system (dashed line). As we see, the mode occupancy in the nonuniform
system deviates substantially from the uniform result and remains below the
maximum occupancy of 1. We note that the nonuniform results corresponding to
cases (1), (2), and (3) of Fig. \ref{figDifferentWidths} almost coincide
with each other and the respective three curves in Fig. \ref{mode-dynamics}%
(b) are almost indistinguishable until the first oscillation maximum. This
implies that the mode occupation dynamics depends only on the shape of the
molecular condensate and not on its size, at least in the short-time limit.
[For example, for a Gaussian shape of the molecular condensate, we find that
the mode occupation dynamics is slightly different.] In the longer time
limit, the first-order correlation function in the fermionic case develops
complicated multipeak structure (similar to the one seen in the momentum
distribution of Fig. \ref{spectrum}). In this case, defining the first-order
coherence length $\Delta k^{(\mathrm{coh})}$ as the half-width at half
maximum becomes less appropriate since this definition ignores the
correlation peaks at large $|k-k^{\prime }|$.

Even though we do not present explicit numerical results for the first-order
correlation function itself, we point out that in the present model the
absolute value of $g_{jj}^{(1)}(k,k^{\prime },t)$ is related to the
second-order correlation function $g_{jj}^{(2)}(k,k^{\prime },t)$ via%
\begin{equation}
|g_{jj}^{(1)}(k,k^{\prime },t)|=\sqrt{|g_{jj}^{(2)}(k,k^{\prime },t)-1|}.
\end{equation}

We now turn to the analysis of the second-order correlation functions for
pairs of atoms in the same spin state, $g_{jj}^{(2)}(k,k^{\prime },t)$ ($%
j=1,2$), and in the opposite spin states, $g_{12}^{(2)}(k,k^{\prime },t)$.

\subsection{Atom-atom correlations}

\subsubsection{Distinguishable fermionic or bosonic atoms}

Since the dissociation of diatomic molecules produces pairs of atoms in two
different spin states which fly apart in opposite directions according to
the momentum conservation, we expect strong correlation signal for atom
pairs with nearly equal but opposite momenta. We refer to this type of
correlation as back-to-back (BB) correlation and quantify it via Glauber's
second-order correlation function,

\begin{equation}
g_{12}^{(2)}(k,k^{\prime },t)=\frac{\langle \widehat{a}_{1}^{\dagger }(k,t)%
\widehat{a}_{2}^{\dagger }(k^{\prime },t)\widehat{a}_{2}(k^{\prime },t)%
\widehat{a}_{1}(k,t)\rangle }{\langle \widehat{a}_{1}^{\dagger }(k,t)%
\widehat{a}_{1}(k,t)\rangle \langle \widehat{a}_{2}^{\dagger }(k^{\prime },t)%
\widehat{a}_{2}(k^{\prime },t)\rangle }.  \label{BB}
\end{equation}%
Apart from the normal ordering of the creation and annihilation operators,
the pair correlation function $g_{12}^{(2)}(k,k^{\prime },t)$ describes the
density-density correlation between the momentum components $k$ and $%
k^{\prime }$ of pairs of atoms in the two spin states. The normalization
with respect to the product of individual densities $n_{j}(k,t)=\langle
\widehat{a}_{j}^{\dagger }(k,t)\widehat{a}_{j}(k,t)\rangle $ ensures that
for uncorrelated states $g_{12}^{(2)}(k,k^{\prime },t)=1$. Due to obvious
symmetry considerations, $g_{12}^{(2)}(k,k^{\prime
},t)=g_{21}^{(2)}(k,k^{\prime },t)$.

The second type of correlation expected to be present in the system is
between pairs of atoms in the same spin state, propagating with the nearly
same momenta, $k\simeq k^{\prime }$. This type of correlation, which we
refer to as collinear (CL) correlation, is due to quantum statistical
effects and represents a manifestation of the Hanbury Brown, and Twiss (HBT)
effect \cite{HBT,Yasuda-Shimizu,Aspect-Westbrook-2005-2007}. The CL
correlations are quantified via the following second-order correlation
function:

\begin{equation}
g_{jj}^{(2)}(k,k^{\prime },t)=\frac{\langle \widehat{a}_{j}^{\dagger }(k,t)%
\widehat{a}_{j}^{\dagger }(k^{\prime },t)\widehat{a}_{j}(k^{\prime },t)%
\widehat{a}_{j}(k,t)\rangle }{\langle \widehat{a}_{j}^{\dagger }(k,t)%
\widehat{a}_{j}(k,t)\rangle \langle \widehat{a}_{j}^{\dagger }(k^{\prime },t)%
\widehat{a}_{j}(k^{\prime },t)\rangle }.  \label{CL}
\end{equation}

\begin{figure}[tbp]
\includegraphics[width=7.5cm]{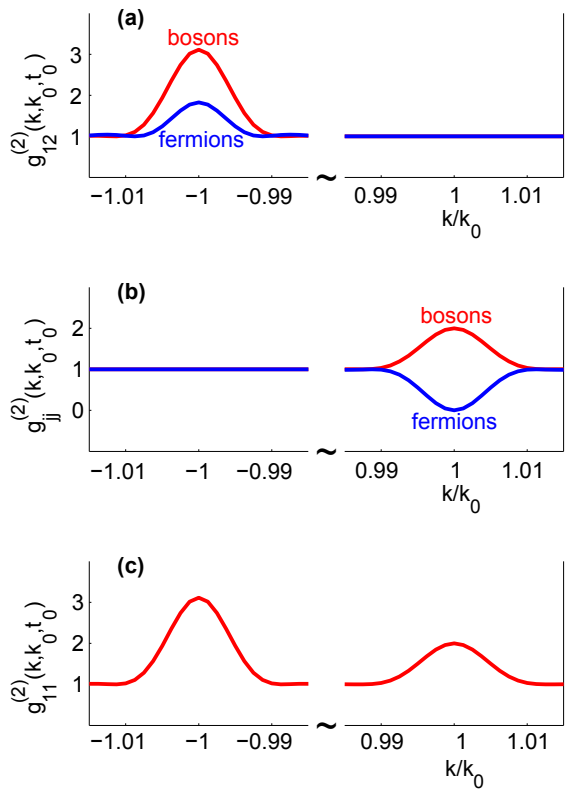} 
\caption{(Color online) Atom-atom pair correlations at $t/t_{0}=1$ for the case of the
molecular BEC profile (1) in Fig. \protect\ref{figDifferentWidths} and
dimensionless detuning $\protect\delta =-9$. (a) $%
g_{12}^{(2)}(k,k_{0},t_{0}) $ as a function of $k$ for fermionic and bosonic
atom pairs in two different spin states, showing the peak corresponding to
strong back-to-back correlation around $k=-k_{0}$. (b) Pair correlation for
the same-spin atoms $g_{11}^{(2)}(k,k_{0},t_{0})$ in dissociation into
distinguishable (fermionic or bosonic) atom pairs, showing a peak (bosons)
or a dip (fermions) at $k=k_{0}$, which correspond, respectively, to bosonic
bunching due to Bose stimulation or fermionic antibunching due to Pauli
blocking. (c) Pair correlation $g_{11}^{(2)}(k,k_{0},t_{0})$ in dissociation
into indistinguishable (same spin state) bosonic atom pairs, showing both
back-to-back and CL correlation signals at $k=-k_{0}$ and $k=k_{0}$. }
\label{g2-a-b-c}
\end{figure}

The linearity of Eqs. (\ref{HeisenbergsEquation}) ensures that one can apply
Wick's theorem to Eqs. (\ref{BB}) and (\ref{CL}) and factorize the
fourth-order operator moments into the sum of products of second-order
moments. Noting in addition that $\langle \widehat{a}_{1}^{\dagger }(k,t)%
\widehat{a}_{2}(k^{\prime },t)\rangle =0$ and $\langle \widehat{a}_{j}(k,t)%
\widehat{a}_{j}(k^{\prime },t)\rangle =0$ in the present model, we obtain
the following results for the BB and CL correlations:%
\begin{gather}
g_{12}^{(2)}(k,k^{\prime },t)=1+\frac{|m_{12}(k,k^{\prime },t)|^{2}}{%
n_{1}(k,t)n_{2}(k^{\prime },t)},  \label{g12} \\
g_{jj}^{(2)}(k,k^{\prime },t)=1\pm \frac{|n_{j}(k,k^{\prime },t)|^{2}}{%
n_{j}(k,t)n_{j}(k^{\prime },t)},  \label{gjj}
\end{gather}%
where the $+$ and $-$ signs stand for bosonic and fermionic atoms,
respectively.

The BB pair correlation $g_{12}^{(2)}(k,k_{0},t=t_{0})$ in which the
momentum of one of the atomic spin components is fixed to the resonant
momentum $k_{0}$, while the momentum $k$ of the opposite spin component is
being varied is plotted in Fig. \ref{g2-a-b-c}(a). The two curves correspond
to the fermionic and bosonic atom statistics, as indicated by the labels. In
both cases, we see a clear correlation peak at $k=-k_{0}$ corresponding to
atom pairs with equal but opposite momenta ($k_{0},-k_{0}$). The width of
the correlation is discussed in Sec. \ref{sec:The-width-of-the-corr}.

The major quantitative difference between the present nonuniform result and
that of a uniform system is that the peak value of the BB correlation $%
g_{12}^{(2)}(-k_{0},k_{0},t)$ becomes smaller than in the uniform system and
that the correlation function acquires a finite width. As has been shown in
Refs. \cite{Fermidiss,PMFT}, the strength of the pair correlation in the
uniform system corresponds to its maximum possible value for a given
occupancy $n_{1,k_{0}}(t)=n_{2,k_{0}}(t)$. More specifically, the pair
correlation upper bound is given by $%
g_{12}^{(2)}(-k_{0},k_{0},t)=1/n_{1,k_{0}}(t)$ in the case of fermionic
atoms and $g_{12}^{(2)}(-k_{0},k_{0},t)=2+1/n_{1,k_{0}}(t)$ in the bosonic
case, whereas $g_{12}^{(2)}(k,k_{0},t)=1$ for any $k\neq -k_{0}$ in both
cases. In other words, the pair correlation in the uniform system is a
Kronecker-like delta function, whereas in the nonuniform case it acquires a
finite width and the peak value is reduced.

\begin{figure}[tbp]
\includegraphics[height=5.2cm]{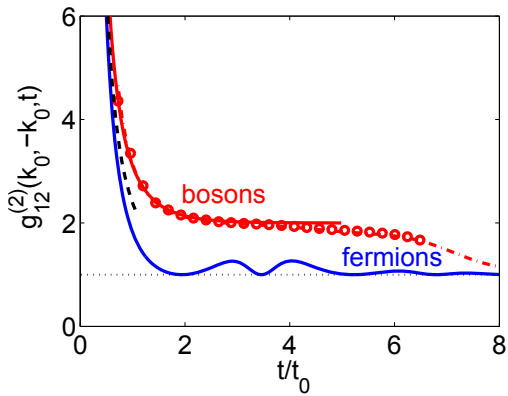} 
\caption{(Color online) Back-to-back pair correlations $%
g_{12}^{(2)}(-k_{0},k_{0},t)$ at equal but opposite peak momenta as a
function of time $t/t_{0}$ for dissociation into fermionic (blue solid
curve) and bosonic (red solid curve) atoms in two spin states from the
numerical simulations within the undepleted molecular-field approximation.
The results are for the molecular density profiles corresponding to case (1)
in Fig. \protect\ref{figDifferentWidths}. The dimensionless detuning is $%
\protect\delta =-9$ as in Fig. \protect\ref{total-atom-number}. The dashed
(black) curve is the short-time analytic result of Eq. (\protect\ref%
{g12-peak-1D}) from Sec. \protect\ref{sec:Analytic}; the curve is extended
to time duration of $t/t_{0}\sim 1$, which, strictly speaking, is beyond the
expected validity of the perturbative theory, $t/t_{0}\ll 1$, yet we see a
reasonably good agreement with the numerical results for $t/t_{0}\sim 1$.
The dash-dotted (red) curve is from the truncated Wigner approach, while the
red circles are from the exact positive-$P$ method; both methods take into
account the molecular-field depletion (see Sec. \protect\ref%
{sec:The-effects-of-interaction} for details). As we see the result from the
undepleted molecular-field approximation (solid red curve, bosons) are in
excellent agreement with the positive-$P$ and truncated Wigner results for
durations up to $t/t_{0}\sim 5$. }
\label{peak-correlation}
\end{figure}

Given that $n_{j,k_{0}}(t)$ in the uniform system is an oscillatory function
for fermions and can reach zero values at certain times, the respective BB
correlation $g_{12}^{(2)}(-k_{0},k_{0},t)=1/n_{1,k_{0}}(t)$ is discontinuous
[$g_{12}^{(2)}(-k_{0},k_{0},t)\rightarrow \infty $] whenever $%
n_{j,k_{0}}(t)\rightarrow 0$. In contrast to this unphysical result, the
peak value of the BB correlation in the nonuniform system is always
continuous and is shown in Fig. \ref{peak-correlation}. The oscillatorylike
structure of the BB correlation for fermions comes from the oscillatory
behavior of the atomic density $n_{j}(k,t)$ and the absolute value of the
anomalous density $|m_{12}(k_{0},-k_{0},t)|$. The oscillations in $%
|m_{12}(k_{0},-k_{0},t)|$ resemble those for a uniform system \cite%
{Fermidiss}, with $%
|m_{k_{0},-k_{0}}(t)|^{2}=n_{j,k_{0}}(t)[1-n_{j,k_{0}}(t)] $, except that
here they do not reach the minimum and maximum values corresponding to
perfect harmonic oscillations in $n_{j,k_{0}}(t)$. The BB correlation for
the case of bosonic atoms is also shown on the same figure (solid red curve)
and does not display any oscillations. We note that the bosonic curve should
have been stopped at $t/t_{0}\sim 3$ as the undepleted molecular
approximation breaks down beyond $t/t_{0}\gtrsim 3$ (the total number of
atoms produced beyond this point in time corresponds to a conversion of more
than 10\% of the initial number of molecules), while this is not the case
for fermionic atoms. We extend the bosonic curve to $t/t_{0}\sim 5$ in order
to make visible its discrepancy with the numerical results based on the
exact positive-$P$ representation (red circles) and the truncated Wigner
method (red dash-dotted curve).

The CL correlation function $g_{11}^{(2)}(k,k_{0},t)$ [with $%
g_{11}^{(2)}(k,k_{0},t)=g_{22}^{(2)}(k,k_{0},t)$] at $t=t_{0}$ is plotted in
Fig. \ref{g2-a-b-c}(b) as a function of $k$. In the case of bosonic atoms we
see the expected HBT peak at $k=k_{0}$, with $g_{11}^{(2)}(k_{0},k_{0},t)=2$
due to bosonic stimulation. In the fermionic case, on the other hand, we see
a dip $g_{11}^{(2)}(k_{0},k_{0},t)=0$ corresponding to fermionic
antibunching due to Pauli blocking. The physics behind these effects is the
same as the CL correlations between the $s$-wave scattered atoms in a
collision of two condensates of metastable $^{4}$He$^{\ast }$ atoms as
observed in Ref. \cite{Perrin2007} and the local correlations in a thermal
cloud of bosonic $^{4}$He$^{\ast }\ $atoms and an ultracold cloud of
fermionic $^{3}$He$^{\ast }$ atoms as observed in Refs. \cite%
{Aspect-Westbrook-2005-2007} (see also \cite{Yasuda-Shimizu}).

\subsubsection{Indistinguishable bosonic atoms}

In the case of dissociation of molecules made of bosonic atom pairs in the
same spin state, the atom-atom pair correlation function is given by%
\begin{gather}
g_{11}^{(2)}(k,k^{\prime },t)=\frac{\langle \widehat{a}_{1}^{\dagger }(k,t)%
\widehat{a}_{1}^{\dagger }(k^{\prime },t)\widehat{a}_{1}(k^{\prime },t)%
\widehat{a}_{1}(k,t)\rangle }{\langle \widehat{a}_{1}^{\dagger }(k,t)%
\widehat{a}_{1}(k,t)\rangle \langle \widehat{a}_{1}^{\dagger }(k^{\prime },t)%
\widehat{a}_{1}(k^{\prime },t)\rangle }  \notag \\
=1+\frac{|n_{1}(k,k^{\prime },t)|^{2}+|m_{11}(k,k^{\prime },t)|^{2}}{%
n_{1}(k,t)n_{1}(k^{\prime },t)}.  \label{g_11Indist}
\end{gather}%
Here, the BB correlation signal comes from the anomalous density term $%
|m_{11}(k,k^{\prime },t)|^{2}$ as it is nonzero only for pairs of momenta $k$
and $k^{\prime }$ that are nearly opposite, whereas the normal density term $%
|n_{1}(k,k^{\prime },t)|^{2}$ is vanishingly small for opposite pairs of
momenta. On the other hand, the CL correlation signal comes from the normal
density term $|n_{1}(k,k^{\prime },t)|^{2}$ which is nonzero for pairs of
nearby momenta, while the anomalous density is vanishingly small when $%
k\simeq k^{\prime }$.

In Fig. \ref{g2-a-b-c}(c) we plot the pair correlation $%
g_{11}^{(2)}(k,k_{0},t_{0})$ as a function of $k$, and we see the
simultaneous presence of two peaks: one at $k=-k_{0}$ representing the BB
correlation due to the momentum conserving pair-production process and the
second peak at $k=k_{0}$ corresponding to the HBT effect.

\subsection{Width of the correlation functions}

\label{sec:The-width-of-the-corr}

An important observable in the experiments on atom-atom correlations is the
width of the correlation functions. Here we discuss the BB and CL
correlation widths during relatively short durations of dissociation,
corresponding to the range of validity of the undepleted molecular-field
approximation.

\begin{figure}[tbp]
\includegraphics[width=8.7cm]{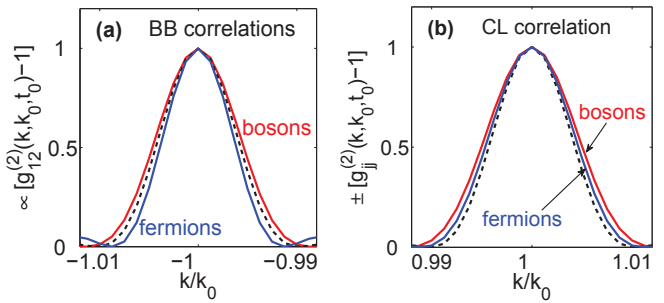}
\caption{(Color online) (a) Back-to-back pair correlation functions, $%
g_{12}^{(2)}(k,k_{0},t_{0})$, at time $t=t_{0}$ for fermions (blue solid
curve) and bosons (red solid curve), corresponding to the molecular BEC
profile (1) of Fig. \protect\ref{figDifferentWidths}. The magnitudes of the
correlation functions are scaled to 1 in order to simplify the comparison of
the correlation widths and the width of the source (dashed curve), that is, the
momentum distribution of the molecular BEC. (b) CL pair correlation
functions, $g_{jj}^{(2)}(k,k_{0},t_{0})$, for fermions (blue solid curve)
and bosons (red solid curve), for the same parameters as in (a). The
fermionic curve for $g_{jj}^{(2)}(k,k_{0},t_{0})-1$ is inverted to allow for
the comparison of the correlation widths. The dashed curve is the momentum
distribution of the source. }
\label{corr-width}
\end{figure}

In Fig.~\ref{corr-width} we show the numerical results for the BB and CL
correlations functions at time $t=t_{0}$ and compare them with the shape of
the source molecular BEC in momentum space. The results correspond to the
molecular BEC density profile (1) of Fig.~\ref{figDifferentWidths}. We see
that in the short-time limit the width and the overall shape of the
correlation functions for both fermionic and bosonic atoms is determined
essentially by the width and the shape of the momentum distribution of the
source molecular BEC.

\begin{figure}[tbp]
\includegraphics[height=5.2cm]{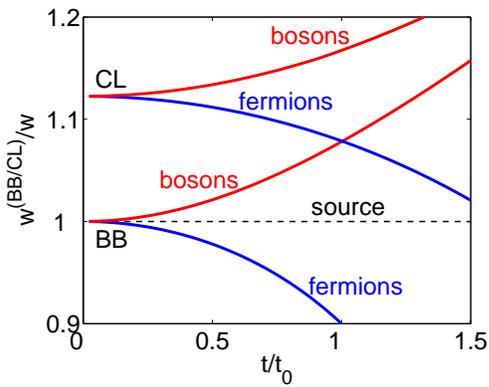} 
\caption{(Color online) Widths (solid lines: bosons, red; fermions, blue)
of the BB and CL pair correlations, $w^{(\mathrm{BB})}$ and $w^{(\mathrm{CL}%
)}$, relative to the width of the source condensate as a function of time,
for the molecular BEC profile (1) of Fig.~ \protect\ref{figDifferentWidths}.}
\label{widths-a-b-c-d}
\end{figure}

For quantitative purposes we define the widths as the half-width at
half maximum and denote them $w^{(\mathrm{BB})}$ and $w^{(\mathrm{CL})}$
for the back-to-back and for CL correlations, respectively. The width
of the source molecular BEC in momentum space is denoted as $w$. The widths
of the BB and CL correlations relative to the source width as a function of
time are shown in Fig.~\ref{widths-a-b-c-d}. The source condensate here
corresponds to the molecular BEC profile (1) of Fig.~\ref{figDifferentWidths}%
, having the momentum width of $w\simeq 1.62/R_{\mathrm{TF}}$, where $R_{%
\mathrm{TF}}$ is the respective TF radius in $x$ space. We see that the BB
correlation width starts from the asymptotic value of $w^{(\mathrm{BB})}=w$,
corresponding to the limit $t\rightarrow 0$, and grows slowly for bosons and
decreases for fermions. A similar behavior is seen in the CL correlation
width, except that it starts from a higher asymptotic value, $w^{(\mathrm{CL}%
)}\simeq 1.12w$, than the BB width. The short-time asymptotic behavior is in
agreement with the analytic results obtained in Sec.~\ref{sec:Analytic}%
.

For longer durations of dissociation---while still within the range of
validity of the undepleted molecular-field approximation---we see that the
BB correlation width grows faster (for bosons) than the CL correlation
width, indicating the possibility that it may eventually become broader than
the CL width. A similar behavior has been recently found in first-principle
simulations of a related system of atomic four-wave mixing via condensate
collisions \cite{Magnus-Karen-4WM}. Moreover, it has been shown that in the
long-time limit the BB correlation width indeed becomes broader than the CL
correlation, in agreement with the experimental measurements of Ref.~\cite%
{Perrin2007}.

For comparison, we have also performed an analysis of the dynamics of the BB
and CL correlation widths for a Gaussian density profile of the source
molecular BEC. The numerical results are in good agreement with the analytic
results presented in Sec. \ref{sec:Analytic-gaussian}.

\subsection{Relative number squeezing}

As an alternative measure of the strength of atom-atom correlations, we now
calculate the variance of relative atom number fluctuations for atoms in
different spin states and with equal but opposite momenta $\pm k_{0}$,
\begin{equation}
V_{k_{0},-k_{0}}(t)=\langle \lbrack \Delta (\widehat{n}_{1,k_{0}}-\widehat{n}%
_{2,-k_{0}})]^{2}\rangle /\Delta _{\mathrm{SN}},  \label{Variance-def}
\end{equation}%
where $\Delta _{\mathrm{SN}}$ is the shot-noise level that originates from
uncorrelated states. The atom number operators are defined by $\widehat{n}%
_{j,\pm k_{0}}(t)=\int_{K}dk\widehat{n}_{j}(k,t)$ [with $\widehat{n}%
_{j}(k,t)=\widehat{a}_{j}^{\dagger }(k,t)\widehat{a}_{j}(k,t)$], where $K$
is the counting length around $\pm k_{0}$. On a computational lattice the
simplest choice of $K$ that does not require explicit binning of the signal
is $K=$ $\Delta k$, where $\Delta k$ is the lattice spacing, and therefore $%
\widehat{n}_{j,\pm k_{0}}(t)=\widehat{n}_{j}(\pm k_{0},t)\Delta k$.

The shot-noise level $\Delta _{\mathrm{SN}}$ is different for bosons and
fermions. For the bosonic case, $\Delta _{\mathrm{SN}}$ is given by the sum
of variances of the individual mode occupancies with Poissonian statistics
(as in the coherent state), implying that $\Delta _{\mathrm{SN}}=\langle
\widehat{n}_{1,k_{0}}\rangle +$ $\langle \widehat{n}_{2,-k_{0}}\rangle $.
For the fermionic case, the sum of the variances of two uncorrelated modes
is $\Delta _{\mathrm{SN}}=\langle \widehat{n}_{1,k_{0}}\rangle (1-\langle
\widehat{n}_{1,k_{0}}\rangle )+$ $\langle \widehat{n}_{2,-k_{0}}\rangle
(1-\langle \widehat{n}_{2,-k_{0}}\rangle )$ \cite{Fermidiss}, which is
independent of the states of individual modes. The relative number variance (\ref{Variance-def}) in both cases can be combined into the following
expression:
\begin{gather}
V_{k_{0},-k_{0}}(t)=1-\frac{\Delta kn_{1}(k_{0},t)}{1-s\Delta kn_{1}(k_{0},t)%
}  \notag \\
\times \lbrack g_{12}^{(2)}(k_{0},-k_{0},t)-g_{11}^{(2)}(k_{0},k_{0},t)-s],
\label{V-raw}
\end{gather}%
where $s=0$ and $s=1$ for bosons and for fermions, respectively, and we have
taken into account that $\langle \widehat{n}_{1,k_{0}}\rangle =\langle
\widehat{n}_{2,-k_{0}}\rangle $ and $%
g_{11}^{(2)}(k_{0},k_{0},t)=g_{22}^{(2)}(-k_{0},-k_{0},t)$. Note that $%
g_{11}^{(2)}(k_{0},k_{0},t)=0$ in the fermionic case due to the Pauli
exclusion principle. Variance $V_{k_{0},-k_{0}}(t)<1$ implies squeezing of
relative number fluctuations below the shot-noise level.

\begin{figure}[tbp]
\includegraphics[height=5.5cm]{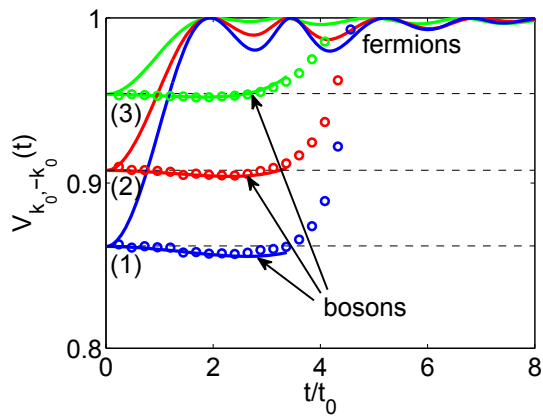} 
\caption{(Color online) Relative number variance for dissociation into
fermionic and bosonic atom pairs (full curves) within the undepleted
molecular approximation. The dashed horizontal lines are the analytic
results in the short-time asymptotic limit [Eq. (\protect\ref%
{variance-analytic})] from Sec. \protect\ref{sec:Analytic}. The three
different sets of curves correspond to three different molecular BEC
profiles of Fig. \protect\ref{figDifferentWidths}. The circles are from the
exact positive-$P$ simulations for bosons that take into account the
molecular depletion. }
\label{variance}
\end{figure}

In Fig. \ref{variance} we plot the relative number variance for dissociation
into fermionic (solid oscillatory curves) and bosonic (solid curves that
initially follow the dashed horizontal lines) atom pairs for the three
different sizes of the molecular BEC density profile of Fig. \ref%
{figDifferentWidths}. The horizontal dashed lines are the respective
analytic results (see Sec. \ref{sec:Analytic}) in the short-time limit
showing a common asymptotic behavior for bosons and fermions in the limit $%
t\rightarrow 0$. In this limit the atomic mode populations are much smaller
than $1$ and the quantum statistical effects do not show up. The numerical
results for bosonic atoms are stopped around $t/t_{0}\sim 3.5$; at this time
the total number of atoms produced corresponds to more than $10$\%
conversion and the undepleted molecular approximation is no longer valid.
For the fermionic atoms the converted fraction remains less than $10$\% for
the entire time window plotted. The circles that follow the bosonic solid
curves for up to $t/t_{0}\sim 3.5$ are from the exact positive-$P$ method
(see Sec. \ref{sec:The-effects-of-interaction}) and demonstrate the
validity of the undepleted molecular approximation within this time window.

We see that both bosonic and fermionic cases display relative number
squeezing that depends strongly on the size of the molecular BEC. The larger
the molecular BEC is, the stronger is the relative number squeezing,
implying that the degrading role of mode-mixing due to the source
inhomogeneity is a weaker effect when the molecular BEC density profile is
closer to uniform. The reduction of squeezing and the nontrivial oscillatory
behavior in the fermionic cases reflect the oscillatory behavior of the
fermionic mode populations and the anomalous density $%
|m_{12}(k_{0},-k_{0},t)|$, as explained in the discussion of the
oscillations in the BB peak correlation in Fig. \ref{peak-correlation}; at
time instances when the populations become close to $0$ or $1$, the
fermionic shot-noise itself becomes vanishingly small and therefore the
degree of squeezing below the shot-noise level diminishes \cite{Fermidiss}.

In all three examples of Fig. \ref{variance} the lattice spacing is chosen
as $\Delta k=1875\simeq \pi /2R_{TF}^{(3)}$ m$^{-1}$, where $R_{TF}^{(3)}$
is the TF radius of the molecular BEC profile (3) of Fig. \ref%
{figDifferentWidths}. This is the largest spacing that is capable of
resolving the relevant details and correlation lengths in momentum space for
all three cases, and we see that the degree of squeezing is not particularly
large in all examples. The strongest squeezing is for the case of the
largest molecular BEC [curve (1)] and is only $\sim 14\%$ at short times.

This relatively modest degrees of squeezing in the examples of Fig. \ref%
{variance} can be contrasted to the ideal situation of $100\%$ squeezing
achievable in a completely uniform system that permits analytic solution in
the undepleted molecular approximation \cite{Fermidiss}. The strong
departure in the actual degree of squeezing from the prediction of the
idealized uniform model is perhaps the strongest manifestation of the role of
mode-mixing in realistic inhomogeneous systems. As shown previously \cite%
{SavageKheruntsyanSpatial,Ogren-directionality}, squeezing can be enhanced
by binning the atomic signal into bins of larger size, in which case Eq. (%
\ref{V-raw}) is no longer applicable.

\section{Analytic treatment in the short-time limit}

\label{sec:Analytic}

In this section we present the results of an analytic treatment of the
problem of molecular dissociation in the short-time limit in 1D, 2D, and 3D.
Our approach is based on perturbative expansion in time, starting with the
operator equations of motion in the undepleted molecular-field
approximation [Eq. (\ref{HeisenbergsEquation})]. Even though the present
results are applicable for even shorter time scales than those of the
numerical treatment of the previous section, their analytic transparency
provides useful insights into the problem of atom-atom correlations as it
has recently been demonstrated for a closely related problem of atomic
four-wave mixing via condensate collisions \cite{Magnus-Karen-4WM}. For
molecular dissociation, the present analytic approach in nonuniform 1D
systems has been employed in Ref. \cite{Magnus-Karen-dissociation-PRA-Rapid}%
; here we present the details of derivations and extend the results to 2D
and 3D systems.

The short-time perturbative treatment is based on the Taylor expansion in
time, up to terms of order $t^{2}$,
\begin{equation}
\widehat{a}_{j}(\mathbf{k},t)=\widehat{a}_{j}(\mathbf{k},0)+\left. t\frac{%
\partial \widehat{a}_{j}(\mathbf{k},t)}{\partial t}\right\vert _{t=0}+\left.
\frac{t^{2}}{2}\frac{\partial ^{2}\widehat{a}_{j}(\mathbf{k},t)}{\partial
t^{2}}\right\vert _{t=0}+\ldots ,  \label{Taylor}
\end{equation}%
which is valid for $t\ll t_{0}$, where $t_{0}\equiv 1/ \chi \sqrt{\rho _{0}}$
is the characteristic time scale. We recall that, in the definition of $%
t_{0} $ for 1D, 2D, and 3D systems, the coupling constant $\chi $ and the
molecular BEC peak density $\rho _{0}$ are to be understood as their 1D, 2D,
and 3D counterparts (see, e.g., \cite{PMFT}), but we suppress the respective
indices for simplicity. In all cases, the units of $\chi $ and $\rho_{0} $
are such that $1/\chi \sqrt{\rho_{0} }$ has units of time. With the preceding
expansion, one can check that the commutation (for bosons) and
anticommutation (for fermions) relations for the creation and annihilation
operators are given by $[\widehat{a}_{i}(\mathbf{k},t),\widehat{a}%
_{j}^{\dagger }(\mathbf{k}^{\prime },t)]_{\mp }\simeq \delta _{ij}\delta (%
\mathbf{k-k}^{\prime })$, up to terms of the order of $t^{2}$.

Using the right-hand sides of the generating equations of motion (\ref%
{HeisenbergsEquation}) for calculating the derivative terms in Eq. (\ref%
{Taylor}), we obtain the following expressions for the anomalous and normal
densities in the lowest order in $t$, in $D=1,2$, and $3$ dimensions:%
\begin{gather}
|m_{12}(\mathbf{k},\mathbf{k}^{\prime },t)|=|\langle \widehat{a}_{1}(\mathbf{%
k},t)\widehat{a}_{2}(\mathbf{k}^{\prime },t)\rangle |  \notag \\
\simeq \frac{t}{(2\pi )^{D/2}}|\widetilde{g}(\mathbf{k}+\mathbf{k}^{\prime
})|  \notag \\
=\frac{t}{(2\pi )^{D}}\left\vert \int d^{D}\mathbf{x}e^{i(\mathbf{k}+\mathbf{%
k}^{\prime })\cdot \mathbf{x}}g(\mathbf{x})\right\vert ,  \label{m12}
\end{gather}%
\begin{gather}
n_{j}(\mathbf{k},\mathbf{k}^{\prime },t)=\langle \widehat{a}_{j}^{\dagger }(%
\mathbf{k},t)\widehat{a}_{j}(\mathbf{k}^{\prime },t)\rangle  \notag \\
\simeq \frac{t^{2}}{(2\pi )^{D}}\int d^{D}\mathbf{q}\widetilde{g}^{*}(%
\mathbf{q}+\mathbf{k})\widetilde{g}(\mathbf{q}+\mathbf{k}^{\prime })  \notag
\\
=\frac{t^{2}}{(2\pi )^{D}}\int d^{D}\mathbf{x}e^{i(\mathbf{k}-\mathbf{k}%
^{\prime })\cdot \mathbf{x}} | g(\mathbf{x}) | ^{2}.  \label{njj}
\end{gather}
Recalling Eqs. (\ref{g12}) and (\ref{gjj}) and viewing Eqs. (\ref{m12}) and (%
\ref{njj}) in the context of BB and CL correlation functions, we see that
the width of the CL correlation between the same-spin atoms [Eq.~(\ref{gjj}%
)] is determined by the square of the Fourier transform of the square of the
effective coupling $g(\mathbf{x})$. The width of the BB correlation [Eq.~(%
\ref{g12})] between the different spin-state atoms, on the other hand, is
determined by the square of the Fourier transform of $g(\mathbf{x})$. Since
the function $g(\mathbf{x})^{2}$ is narrower than $g(\mathbf{x})$ and the
converse is true for their respective Fourier transforms, we immediately
deduce that the CL momentum correlation is generally broader than the BB
correlation. These conclusions are true for any shape of the source
condensate and apply to both bosonic and fermionic statistics in the short-time limit.

\subsection{Thomas-Fermi parabolic density profiles}

In this subsection we consider specific density profiles of the molecular
BEC and use the perturbative results of Eqs. (\ref{m12}) and (\ref{njj}) for
calculating atom-atom pair correlations and the relative number squeezing in
the short-time limit. As one of the most typical situations, we start with a
parabolic density profile characteristic of an interacting BEC in a harmonic
trap in the TF limit.

\subsubsection{One dimension (1D)}

In 1D we assume that the molecular BEC profile is given by an
inverted TF parabola, $\rho_{0}(x)=\rho_{0}(1-x^{2}/R_{TF}^{2})$ for $%
x<R_{TF}$ [and $\rho_{0}(x)=0$ for $x\geq R_{TF}$], where $\rho_{0}\equiv
\rho_{0}(0)$ is the peak density. Accordingly, the effective coupling
constant is given by $g(x)=\chi \sqrt{\rho_{0}}(1-x^{2}/R_{TF}^{2})^{1/2}$.
The integrals appearing in Eqs. (\ref{m12}) and (\ref{njj}) can be expressed
in terms of Bessel functions, using the following integral representation
\cite{Bateman}:
\begin{equation}
J_{\nu }(q)=\frac{2(q/2)^{\nu }}{\sqrt{\pi }\Gamma (\nu +1/2)}%
\int\nolimits_{0}^{1}d\xi \,(1-\xi ^{2})^{\nu -1/2}\cos (q\xi ),
\end{equation}%
where $\nu >-1/2$ and $\Gamma (\nu )$ is the gamma function. Accordingly,
the integrals in Eqs. (\ref{m12}) and (\ref{njj}) yield
\begin{equation}
|m_{12}(k,k^{\prime },t)|\simeq \frac{t\chi \sqrt{\rho _{0}}R_{TF}}{2}\;%
\frac{J_{1}\left( (k+k^{\prime })R_{TF}\right) }{(k+k^{\prime })R_{TF}},
\label{m12-1D}
\end{equation}%
\begin{equation}
n_{j}(k,k^{\prime },t)\simeq \frac{\sqrt{2}t^{2}\chi ^{2}\rho _{0}R_{TF}}{%
\sqrt{\pi }}\;\frac{J_{3/2}\left( (k-k^{\prime })R_{TF}\right) }{\left[
(k-k^{\prime })R_{TF}\right] ^{3/2}}.  \label{njj-1D}
\end{equation}

Substitution of these expressions into Eqs. (\ref{g12}) and (\ref{gjj})
leads to the following results for the BB and CL pair correlation functions:%
\begin{equation}
g_{12}^{(2)}(k,k^{\prime },t)\simeq 1+\frac{9\pi ^{2}}{16t^{2}\chi ^{2}\rho
_{0}}\frac{\left[ J_{1}\left( (k+k^{\prime })R_{TF}\right) \right] ^{2}}{%
\left[ (k+k^{\prime })R_{TF}\right] ^{2}},  \label{g12-1D}
\end{equation}%
\begin{equation}
g_{jj}^{(2)}(k,k^{\prime },t)\simeq 1\pm \frac{9\pi }{2}\frac{\left[
J_{3/2}\left( (k-k^{\prime })R_{TF}\right) \right] ^{2}}{\left[ (k-k^{\prime
})R_{TF}\right] ^{3}}.  \label{gjj-1D}
\end{equation}%
These results are valid for $t\ll t_{0}$, and are plotted in Fig. \ref%
{analytic-corr-1D} as a function of $k$, for $k^{\prime }=k_{0}$ and $%
t=0.1t_{0}$. Once scaled with respect to the corresponding peak values and
plotted as in Fig. \ref{corr-width}, the curves in Fig. \ref%
{analytic-corr-1D} follow closely the bosonic and fermionic numerical
results shown in Fig. \ref{corr-width}. Interestingly, even beyond the
strict range of applicability of the analytic results of Eqs. (\ref{g12-1D})
and (\ref{gjj-1D}), they show good agreement with the numerical results of
the previous section, valid for up to $t\sim t_{0}$. For example, at $%
t=t_{0} $ the differences between the BB and CL correlation widths,
evaluated using the analytic and numerical results, are less than $10\%$.

\begin{figure}[tbp]
\includegraphics[width=8.7cm]{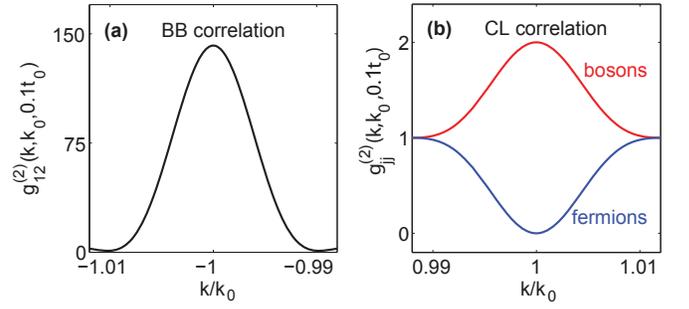}
\caption{(Color online) Analytic results of Eqs. (\protect\ref{g12-1D}) and (\protect\ref{gjj-1D}) for the atom-atom pair correlation functions at $%
t=0.1t_{0}$, for the case of a TF parabolic profile of the molecular BEC
with $R_{TF}=250$ $\protect\mu m$. Other parameters are as in Fig. \protect
\ref{total-atom-number}. (a) $g_{12}^{(2)}(k,k_{0},0.1t_{0})$ as a function
of $k$ for fermionic and bosonic atom pairs in two different spin states,
showing the back-to-back correlation peak at $k=-k_{0}$; (b) pair
correlation for the same-spin atoms $g_{11}^{(2)}(k,k_{0},t)$ in
dissociation into distinguishable (fermionic or bosonic) atom pairs, showing
a CL HBT peak for bosons or a HBT dip for fermions at $k=k_{0}$. }
\label{analytic-corr-1D}
\end{figure}

The correlation widths that follow from Eqs. (\ref{g12}) and (\ref{gjj}) are%
\begin{eqnarray}
w^{(\mathrm{BB})} &=&w\simeq 1.62/R_{\mathrm{TF}}, \\
w^{(\mathrm{CL})} &\simeq &1.12w\simeq 1.81/R_{\mathrm{TF}},
\end{eqnarray}%
and therefore $w^{(\mathrm{CL})}/w^{(\mathrm{BB})}\simeq 1.12$. Here $%
w\simeq 1.62/R_{\mathrm{TF}}$ is the width of the molecular BEC momentum
distribution, $n_{0}(k)=|\int dx\sqrt{\rho_{0}(x)}\exp (-ikx)/(2\pi
)^{1/2}|^{2}$, given by%
\begin{equation}
n_{0}(k)=\frac{\pi \rho _{0}}{2}\frac{\left[ J_{1}(kR_{\mathrm{TF}})\right]
^{2}}{k^{2}}.  \label{1Dmolmomdist}
\end{equation}

Inverting the relationship between the second- and first-order correlation
functions, $|g_{jj}^{(1)}(k,k^{\prime },t)|=\sqrt{|g_{jj}^{(2)}(k,k^{\prime
},t)-1|}$, we can also find the width of the first-order correlation
function, $\Delta k^{(\mathrm{coh})}$, which gives the phase coherence
length and defines the size of an atomic mode in free space. From the above
analytic result for $g_{jj}^{(2)}(k,k^{\prime },t),$ we find that the
first-order coherence length in the short-time limit is approximately $%
\Delta k^{(\mathrm{coh})}=2.50/R_{\mathrm{TF}}$. In terms of the momentum
width of the source condensate $w$, this corresponds to $\Delta k^{(\mathrm{%
coh})}=1.54w$, and shows that the first-order coherence length is larger
than the second-order correlation length $w^{(\mathrm{CL})}$.

Using the asymptotic behavior of the Bessel function, $J_{\nu }(z)\simeq
(z/2)^{\nu }/\Gamma (\nu +1)$ for $z\ll 1$ ($\nu \neq -1,-2,\ldots $), Eqs. (\ref{m12-1D}) and (\ref{njj-1D}) at $k^{\prime }=\mp k$ give, respectively,
the peak value of the anomalous density $m_{12}(k,t)\equiv m_{12}(k,-k,t)$
and the atomic momentum distribution in the spin state $j$, $%
n_{j}(k,t)\equiv n_{j}(k,k,t)$:
\begin{gather}
|m_{12}(k,t)|\simeq \frac{t\chi \sqrt{\rho _{0}}R_{\mathrm{TF}}}{4},
\label{m1D} \\
n_{j}(k,t)\simeq \frac{2t^{2}\chi ^{2}\rho _{0}R_{\mathrm{TF}}}{3\pi }.
\label{n1D}
\end{gather}%
The peak densities are uniform in the short-time limit, corresponding to
\emph{spontaneous} initiation of dissociation, which populates the atomic
modes uniformly without the need to strictly conserve energy. A
double-peaked structure with maxima around $|k|=\pm k_{0}$ (which in 2D and
3D turns into a circle and a sphere of radius $|\mathbf{k}|=k_{0}$ \cite%
{Fermidiss,Savage}) forms later in time \cite{twinbeams}, when the present
analytic results are not applicable.

Similarly, we can find that the peak values of the pair correlation
functions (\ref{g12-1D}) and (\ref{gjj-1D}) are given by%
\begin{equation}
g_{12}^{(2)}(k,-k,t)\simeq 1+\frac{9\pi ^{2}}{64t^{2}\chi ^{2}\rho _{0}},
\label{g12-peak-1D}
\end{equation}%
\begin{equation}
g_{jj}^{(2)}(k,k,t)=\left\{
\begin{array}{l}
2,\text{ bosons,} \\
0,\text{ fermions.}%
\end{array}%
\right.  \label{gjj-peak}
\end{equation}%
The inverse square dependence of the BB correlation peak, $%
g_{12}^{(2)}(k,-k,t)$, on time (for $t\ll t_{0}$) is in good agreement with
the numerical results of the undepleted molecular-field approximation, as
well as with the results of first-principles simulations using the positive-$%
P$ method (see Fig. \ref{peak-correlation}). The peak values of the CL
correlation, $g_{jj}^{(2)}(k,k,t)$, correspond to the expected HBT bunching
for bosons and the HBT dip for fermions.

Finally we calculate the short-time asymptote for the relative number
variance [Eq. (\ref{V-raw})] and find the following simple result:%
\begin{equation}
V_{k_{0},-k_{0}}(t)\simeq 1-\frac{3\pi \Delta kR_{\mathrm{TF}}}{32}.
\label{variance-analytic}
\end{equation}%
In this limit the atomic mode populations are much smaller than 1, bosonic
and fermionic shot noises are approximately equal to each other, and we do
not see any difference in the relative number squeezing for bosons and
fermions. In Fig. \ref{variance}, the results of Eq.\ (\ref%
{variance-analytic}) for three different sizes of the initial molecular BEC
are shown as horizontal dashed lines. As we see, the relative number
squeezing at short times depends merely on the size of the source condensate
$R_{\mathrm{TF}}$ and the size of the counting cell $\Delta k$. The small
geometric prefactor in the second term is determined by the shape of the
source condensate (TF parabola in the present case). Together with the
resolution requirement of $\Delta k\lesssim 1/R_{\mathrm{TF}}$, the
smallness of this prefactor ensures that $V_{k_{0},-k_{0}}(t)>0$; at the
same time this implies that the raw (unbinned) squeezing can be very weak
for small $R_{\mathrm{TF}}$.

\subsubsection{Two dimensions (2D)}

In 2D, the TF density profile of the molecular BEC in a harmonic
trap is given by $\rho_{0} (\mathbf{x})=\rho _{0}(1-x^{2}/R_{\mathrm{TF}%
,x}^{2}-y^{2}/R_{\mathrm{TF},y}^{2})$ for $x^{2}/R_{\mathrm{TF}%
,x}^{2}+y^{2}/R_{\mathrm{TF},y}^{2}<1$ [and $\rho_{0} (\mathbf{x})=0$
otherwise], and therefore $g(\mathbf{x})=\chi \sqrt{\rho _{0}}(1-x^{2}/R_{%
\mathrm{TF},x}^{2}-y^{2}/R_{\mathrm{TF},y}^{2})^{1/2}$. The analysis of the
short-time asymptotic behavior of the correlation functions is essentially
the same as in 1D, except that one has to specify in advance the direction
of the displacement $\Delta \mathbf{k}$ between the pair of momentum vectors
$\mathbf{k}$ and $\mathbf{k}^{\prime }$ for which the correlations are being
analyzed. For definiteness, we consider BB and CL correlations for which the
displacement $\Delta \mathbf{k}$ is along one of the Cartesian coordinates, $%
k_{i}$ ($i=x,y$). In other words, we consider correlations between $\mathbf{k}
$ and $\mathbf{k}^{\prime }=\pm \mathbf{k+e}_{i}\Delta k_{i}$, where $%
\mathbf{e}_{i}$ is the unit vector in the $k_{i}$ direction; the plus sign
is for the CL correlation, and the minus sign is for the BB correlation. In
the results that follow, the dependence on $\mathbf{k}$ is absent and we
omit it for notational simplicity.

The 2D integrals in Eqs. (\ref{m12}) and (\ref{njj}) can again be performed
in terms of Bessel functions (see Appendix \ref{2D-integrals}), and we obtain%
\begin{equation}
|m_{12}(k_{i},k_{i}^{\prime },t)|\simeq \frac{t\chi \sqrt{\rho _{0}}%
\overline{R_{\mathrm{TF}}}^{2}}{2\sqrt{2\pi }}\;\frac{J_{3/2}\left(
(k_{i}+k_{i}^{\prime })R_{\mathrm{TF},i}\right) }{\left[ (k_{i}+k_{i}^{%
\prime })R_{\mathrm{TF},i}\right] ^{3/2}},  \label{m12-2D}
\end{equation}%
\begin{equation}
n_{j}(k_{i},k_{i}^{\prime },t)\simeq \frac{t^{2}\chi ^{2}\rho _{0}\overline{%
R_{\mathrm{TF}}}^{2}}{\pi }\frac{J_{2}\left( (k_{i}-k_{i}^{\prime })R_{%
\mathrm{TF},i}\right) }{\left[ (k_{i}-k_{i}^{\prime })R_{\mathrm{TF},i}%
\right] ^{2}},  \label{njj-2D}
\end{equation}%
where $\overline{R_{\mathrm{TF}}}=(R_{\mathrm{TF},x}R_{\mathrm{TF},y})^{1/2}$
is the geometric mean TF radius. The preceding expressions lead to the following
explicit results for the second-order correlation functions:%
\begin{equation}
g_{12}^{(2)}(k_{i},k_{i}^{\prime },t)\simeq 1+\frac{8\pi }{t^{2}\chi
^{2}\rho _{0}}\frac{\left[ J_{3/2}\left( (k_{i}+k_{i}^{\prime })R_{\mathrm{TF%
},i}\right) \right] ^{2}}{\left[ (k_{i}+k_{i}^{\prime })R_{\mathrm{TF},i}%
\right] ^{3}},  \label{g12-2D}
\end{equation}%
\begin{equation}
g_{jj}^{(2)}(k_{i},k_{i}^{\prime },t)\simeq 1\pm \frac{64\left[ J_{2}\left(
(k_{i}-k_{i}^{\prime })R_{\mathrm{TF},i}\right) \right] ^{2}}{\left[
(k_{i}-k_{i}^{\prime })R_{\mathrm{TF},i}\right] ^{4}}.  \label{gjj-2D}
\end{equation}

The dependence of $g_{12}^{(2)}(k_{i},k_{i}^{\prime },t)$ on $%
k_{i}+k_{i}^{\prime }$ and its peak at $k_{i}+k_{i}^{\prime }=0$ implies
that it describes the BB correlation of different spin atoms with equal but
opposite momenta, for which $\mathbf{k}^{\prime }\simeq -\mathbf{k}$ and is
offset from $-\mathbf{k}$ by an amount $\Delta k_{i}$ in the $i$th
direction. Similarly, the dependence of $g_{jj}^{(2)}(k_{i},k_{i}^{\prime
},t)$ on $k_{i}-k_{i}^{\prime }$ and its peak at $k_{i}-k_{i}^{\prime }=0$
corresponds to the CL correlation between pairs of atoms in the same spin
state, for which $\mathbf{k}^{\prime }\simeq \mathbf{k}$ and is offset by an
amount $\Delta k_{i}$. The qualitative behavior of the BB and CL correlation
functions is the same as in 1D (Fig. \ref{analytic-corr-1D}) whereas the
quantitative differences enter through the width of the correlations and
their respective peak values.

The widths of the BB and CL correlation functions in 2D are%
\begin{eqnarray}
w_{i}^{(\mathrm{BB})} &=&w_{i}\simeq 1.81/R_{\mathrm{TF},i}, \\
w_{i}^{(\mathrm{CL})} &\simeq &1.10w_{i}\simeq 1.99/R_{\mathrm{TF},i},
\end{eqnarray}%
and therefore $w_{i}^{(\mathrm{CL})}/w_{i}^{(\mathrm{BB})}\simeq 1.10$.
Relative to the source width $w_{i}$, the CL correlation width in 2D is
narrower than in 1D, while the BB correlation width is equal to $w_{i}$.
Here, $w_{i}\simeq 1.81/R_{\mathrm{TF},i}$ is the width of the momentum
distribution of the molecular BEC along $i$, found from $n_{0}(k_{i})=|\int
d^{2}\mathbf{x}\sqrt{\rho_{0} (\mathbf{x})}\exp (-ik_{i}x_{i})/(2\pi )|^{2}$
(see Appendix \ref{2D-integrals}) as
\begin{equation}
n_{0}(k_{i})=\frac{\pi \rho _{0}\overline{R_{\mathrm{TF}}}^{4}}{2}\frac{%
\left[ J_{3/2}(k_{i}R_{\mathrm{TF},i})\right] ^{2}}{( k_{i}R_{\mathrm{TF}%
,i}) ^{3}}.  \label{2Dmolmomdist}
\end{equation}

The first-order coherence length in 2D, which follows from $%
|g_{jj}^{(1)}(k_{i},k_{i}^{\prime },t)|=\sqrt{|g_{jj}^{(2)}(k_{i},k_{i}^{%
\prime },t)-1|}$, is given by $\Delta k_{i}^{(\mathrm{coh})}\simeq 2.75/R_{%
\mathrm{TF,}i}$ and therefore $\Delta k_{i}^{(\mathrm{coh})}\simeq 1.52w_{i}$%
.

The peak values of the anomalous and normal densities, $m_{12}(\mathbf{k}%
,t)\equiv m_{12}(\mathbf{k},-\mathbf{k},t)$ and $n_{j}(\mathbf{k},t)\equiv
n_{j}(\mathbf{k,k},t)$, are given by
\begin{gather}
m_{12}(\mathbf{k},t)\simeq \frac{t\chi \sqrt{\rho _{0}}\overline{R_{\mathrm{%
TF}}}^{2}}{6\pi }, \\
n_{j}(\mathbf{k},t)\simeq \frac{t^{2}\chi ^{2}\rho _{0}\overline{R_{\mathrm{%
TF}}}^{2}}{8\pi }.
\end{gather}

The peak CL correlation function $g_{jj}^{(2)}(\mathbf{k},\mathbf{k},t)$ is
the same as in Eq. (\ref{gjj-peak}), whereas the peak BB correlation is
given by
\begin{equation}
g_{12}^{(2)}(\mathbf{k},-\mathbf{k},t)\simeq 1+\frac{16}{9t^{2}\chi ^{2}\rho
_{0}}.  \label{g12-peak-2D}
\end{equation}%
This result is qualitatively similar to the 1D result of Eq. (\ref%
{g12-peak-1D}), except that the numerical prefactor in the second term is
different. The same is true for the relative number variance. In 2D it is
given by
\begin{equation}
V_{\mathbf{k}_{0},\mathbf{-k}_{0}}(t)\simeq 1-\frac{2(\overline{\Delta k}%
)^{2}(\overline{R_{\mathrm{TF}}})^{2}}{9\pi },
\end{equation}%
where $\overline{\Delta k}=(\Delta k_{x}\Delta k_{y})^{1/2}$ is the
geometric mean lattice spacing, with $\overline{\Delta k}^{2}$ giving the
counting area. Comparing this result with $V_{k_{0},-k_{0}}(t)$ in 1D [Eq. (%
\ref{variance-analytic})], we see that the raw squeezing is weaker in 2D than
in 1D, for the same size of the molecular BEC and the same lattice spacing.

\subsubsection{Three dimensions (3D)}

In 3D the TF density profile is given by $\rho_{0} (\mathbf{x})=\rho
_{0}(1-x^{2}/R_{\mathrm{TF},x}^{2}-y^{2}/R_{\mathrm{TF},y}^{2}-z^{2}/R_{%
\mathrm{TF},z}^{2})$ for $x^{2}/R_{\mathrm{TF},x}^{2}+y^{2}/R_{\mathrm{TF}%
,y}^{2}+z^{2}/R_{\mathrm{TF},z}^{2}<1$ [and $\rho_{0} (\mathbf{x})=0$
otherwise], and therefore $g(\mathbf{x})=\chi \sqrt{\rho _{0}}(1-x^{2}/R_{%
\mathrm{TF},x}^{2}-y^{2}/R_{\mathrm{TF},y}^{2}-z^{2}/R_{\mathrm{TF}%
,z}^{2})^{1/2}$. As in 2D, we are interested in BB and CL density
correlations between two momentum components at $\mathbf{k}$ and $\mathbf{k}%
^{\prime }$, for which the displacement $\Delta \mathbf{k=k-k}^{\prime }$ is
along one of the Cartesian coordinates, $k_{i}$, where $i=x,y,z$. The 3D
integrals in Eqs. (\ref{m12}) and (\ref{njj}) can again be performed in
terms of Bessel functions (see Appendix \ref{3D-integrals}), and we obtain%
\begin{equation}
|m_{12}(k_{i},k_{i}^{\prime },t)| \simeq \frac{t\chi \sqrt{\rho _{0}}%
\overline{R_{\mathrm{TF}}}^{3}}{4\pi }\frac{J_{2}\left( (k_{i}+k_{i}^{\prime
})R_{\mathrm{TF},i}\right) }{\left[ (k_{i}+k_{i}^{\prime })R_{\mathrm{TF},i}%
\right] ^{2}},  \label{m12-3D}
\end{equation}%
\begin{equation}
n_{j}(k_{i},k_{i}^{\prime },t)\simeq \frac{t^{2}\chi ^{2}\rho _{0}\overline{%
R_{\mathrm{TF}}}^{3}}{\pi \sqrt{2\pi }}\;\frac{J_{5/2}\left(
(k_{i}-k_{i}^{\prime })R_{\mathrm{TF},i}\right) }{\left[ (k_{i}-k_{i}^{%
\prime })R_{\mathrm{TF},i}\right] ^{5/2}},  \label{njj-3D}
\end{equation}%
where $\overline{R_{\mathrm{TF}}}=(R_{\mathrm{TF},x}R_{\mathrm{TF},y}R_{%
\mathrm{TF},z})^{1/3}$ is the geometric mean TF radius. The BB and CL
correlations following from these expressions are%
\begin{equation}
g_{12}^{(2)}(k_{i},k_{i}^{\prime },t)\simeq 1+\frac{225\pi ^{2}}{16t^{2}\chi
^{2}\rho _{0}}\frac{\left[ J_{2}\left( (k_{i}+k_{i}^{\prime })R_{\mathrm{TF}%
,i}\right) \right] ^{2}}{\left[ (k_{i}+k_{i}^{\prime })R_{\mathrm{TF},i}%
\right] ^{4}},
\end{equation}%
\begin{equation}
g_{jj}^{(2)}(k_{i},k_{i}^{\prime },t)\simeq 1\pm \frac{225\pi }{2}\frac{%
\left[ J_{5/2}\left( (k_{i}-k_{i}^{\prime })R_{\mathrm{TF},i}\right) \right]
^{2}}{\left[ (k_{i}-k_{i}^{\prime })R_{\mathrm{TF},i}\right] ^{5}}.
\end{equation}%
As in 2D, the qualitative behavior of the BB and CL correlation functions is
the same as in 1D (Fig. \ref{analytic-corr-1D}), whereas the quantitative
differences enter through the width and the peak values.

The widths of the BB and CL correlations in 3D are given by
\begin{eqnarray}
w_{i}^{(\mathrm{BB})} &=&w_{i}\simeq 1.99/R_{\mathrm{TF},i}, \\
w_{i}^{(\mathrm{CL})} &\simeq &1.08w_{i}\simeq 2.16/R_{\mathrm{TF},i},
\end{eqnarray}%
and therefore $w_{i}^{(\mathrm{CL})}/w_{i}^{(\mathrm{BB})}\simeq 1.08$. The
CL correlation width in 3D relative to the source width $w_{i}$ is smaller
than in 1D and 2D, whereas the relative BB correlation width is the same.
Here $w_{i}\simeq 1.99/R_{\mathrm{TF},i}$ is the width of the momentum
distribution of the molecular BEC along $i$, found from $n_{0}(k_{i})=|\int
d^{3}\mathbf{x}\sqrt{\rho_{0} (\mathbf{x})}\exp (-ik_{i}x_{i})/(2\pi
)^{3/2}|^{2}$ (see Appendix \ref{3D-integrals}) as%
\begin{equation}
n_{0}(k_{i})=\frac{\pi \rho _{0}\overline{R_{\mathrm{TF}}}}{2}^{6}\frac{%
\left[ J_{2}(k_{i}R_{\mathrm{TF},i})\right] ^{2}}{(k_{i}R_{\mathrm{TF}%
,i})^{4}}.  \label{3Dmolmomdist}
\end{equation}

The first-order coherence length in 3D, following from $%
|g_{jj}^{(1)}(k_{i},k_{i}^{\prime },t)|=\sqrt{|g_{jj}^{(2)}(k_{i},k_{i}^{%
\prime },t)-1|}$, is given by $\Delta k_{i}^{(\mathrm{coh})}\simeq 2.99/R_{%
\mathrm{TF,}i}$ and therefore $\Delta k_{i}^{(\mathrm{coh})}\simeq 1.50w_{i}$%
. This is again larger that the second-order CL correlation width $w_{i}^{(%
\mathrm{CL})}\simeq 1.08w_{i}$.

The peak values of the anomalous and normal densities in 3D are given by%
\begin{eqnarray}
n_{j}(\mathbf{k},t) &\simeq &\frac{t^{2}\chi ^{2}\rho _{0}\overline{R_{%
\mathrm{TF}}}^{3}}{15\pi ^{2}}, \\
m_{12}(\mathbf{k},t) &\simeq &\frac{t\chi \sqrt{\rho _{0}}\overline{R_{%
\mathrm{TF}}}^{3}}{32\pi },
\end{eqnarray}%
whereas the peak BB correlation is%
\begin{equation}
g_{12}^{(2)}(\mathbf{k},-\mathbf{k},t)\simeq 1+\frac{15^{2}\pi ^{2}}{%
32^{2}t^{2}\chi ^{2}\rho _{0}}.
\end{equation}%
The peak CL correlation function $g_{jj}^{(2)}(\mathbf{k},\mathbf{k},t)$ is
the same as in Eq. (\ref{gjj-peak}). The result for $g_{12}^{(2)}(\mathbf{k}%
,-\mathbf{k},t)$ is qualitatively similar to the 1D and 2D results of Eqs. (\ref{g12-peak-1D}) and (\ref{g12-peak-2D}), except that the numerical
prefactor in the second term is different.

Finally, the relative number squeezing in 3D is determined by%
\begin{equation}
V_{\mathbf{k}_{0},\mathbf{-k}_{0}}(t)\simeq 1-\frac{15(\overline{\Delta k}%
)^{3}(\overline{R_{\mathrm{TF}}})^{3}}{32^{2}},
\end{equation}%
where $\overline{\Delta k}=(\Delta k_{x}\Delta k_{y}\Delta k_{z})^{1/3}$ is
the geometric mean lattice spacing, with $\overline{\Delta k}^{3}$ giving
the counting volume. The raw squeezing in 3D is weaker than in 1D and 2D,
for the same size of the molecular BEC and the same lattice spacing.

\subsection{Gaussian density profiles}

\label{sec:Analytic-gaussian}

In this subsection we present the short-time analytic results for the
correlation functions in the case of a Gaussian density profile of the
source molecular BEC, $\rho _{0}(\mathbf{x})=\rho _{0}\exp
(-\sum\nolimits_{i=1}^{D}x_{i}^{2}/2S_{i}^{2})$, where $\rho _{0}$ is the
peak density and $S_{i}$ is the rms width in the $i$th direction. The
results for 1D, 2D, and 3D systems can be combined through setting $D=1,$ $2,$
or $3$. As in the case of the TF density profile, we are interested in BB
and CL density correlations between two momentum components at $\mathbf{k}$
and $\mathbf{k}^{\prime }$, for which the displacement $\Delta \mathbf{k=k-k}%
^{\prime }$ is along one of the Cartesian coordinates, $k_{i}$, where $%
i=x,y,z$ in 3D (in 2D, $i=x,y$, while in 1D, $i=x$). We define the momentum
width of the molecular condensate along $k_{i}$ via $\sigma _{i}$, which
corresponds to the rms width $\sigma _{i}=1/2S_{i}$ of a Gaussian%
\begin{equation}
n_{0}(k_{i})=\frac{\rho _{0}}{2^{D}(\overline{\sigma })^{2D}}\exp
(-k_{i}^{2}/2\sigma _{i}^{2}),
\end{equation}%
where $n_{0}(k_{i})$ is defined according to $n_{0}(k_{i})=|\int d^{D}%
\mathbf{x}\sqrt{\rho_{0} (\mathbf{x})}\exp (-ik_{i}x_{i})/(2\pi )^{D/2}|^{2}$
and $\overline{\sigma }=\left( \prod\nolimits_{i=1}^{D}\sigma _{i}\right)
^{1/D}$ is the geometric mean width.

With these definitions, the integrals in Eqs. (\ref{m12}) and (\ref{njj}),
with $g(\mathbf{x})=\chi \sqrt{\rho _{0}(\mathbf{x})}$, give
\begin{gather}
|m_{12}(k_{i},k_{i}^{\prime },t)|\simeq \frac{t\chi \sqrt{\rho _{0}}}{%
2^{D}\pi ^{D/2}\overline{\sigma }}\exp \left[ -(k_{i}+k_{i}^{\prime
})^{2}/4\sigma _{i}^{2}\right] ,  \label{m1D_Gaussian} \\
n_{j}(k_{i},k_{i}^{\prime },t)\simeq \frac{t^{2}\chi ^{2}\rho _{0}}{%
2^{D+D/2}\pi ^{D/2}\overline{\sigma }}\exp \left[ -(k_{i}-k_{i}^{\prime
})^{2}/8\sigma _{i}^{2}\right],  \label{n1D_Gaussian}
\end{gather}%
and therefore the BB and CL correlation functions are%
\begin{equation}
g_{12}^{(2)}(k,k^{\prime },t)\simeq 1+\frac{2^{D}}{t^{2}\chi ^{2}\rho _{0}}%
\exp \left[ -(k_{i}+k_{i}^{\prime })^{2}/2\sigma _{i}^{2}\right] ,
\end{equation}%
\begin{equation}
g_{jj}^{(2)}(k_{i},k_{i}^{\prime },t)\simeq 1\pm \exp \left[
-(k_{i}-k_{i}^{\prime })^{2}/4\sigma _{i}^{2}\right] .  \label{gjjGaussian}
\end{equation}

The BB and CL correlation widths are%
\begin{eqnarray}
\sigma ^{(\mathrm{BB})} &=&\sigma _{i}=1/2S_{i}, \\
\sigma ^{(\mathrm{CL})} &=&\sqrt{2}\sigma _{i}=1/\sqrt{2}S_{i}.
\end{eqnarray}%
The CL correlation width, relative to the source width $\sigma _{i}$, in the
present Gaussian case is broader than in the case of a TF parabolic density
profile of the source molecular BEC, whereas the BB correlation width is
equal to the width of the source as before.

The relative number squeezing is given by%
\begin{equation}
V_{\mathbf{k}_{0},-\mathbf{k}_{0}}(t)\simeq 1-\left( \frac{2}{\pi }\right)
^{D/2}(\overline{\Delta k})^{D}(\overline{S})^{D},
\end{equation}%
where $\overline{S}=\left( \prod\nolimits_{i=1}^{D}S_{i}\right) ^{1/D}$ is
the geometric mean width of the molecular BEC in coordinate space and $%
\overline{\Delta k}=\left( \prod\nolimits_{i=1}^{D}\Delta k_{i}\right)
^{1/D} $. The general qualitative conclusions about relative number
squeezing in 1D, 2D, and 3D remain the same as for the TF parabolic density
profile of the source molecular BEC.

\section{Effects of molecular depletion and collisional interactions}

\label{sec:The-effects-of-interaction}

In this section we discuss the role of molecular depletion and $s$-wave
scattering interactions. We treat these effects explicitly using two
alternative phase-space representation techniques: first-principles
simulations using the positive-$P$ method \cite{Drummond-Gardiner} and a
truncated Wigner approximation \cite{Steel,Castin}. Owing to the fact that
these phase-space methods are currently well established for bosonic fields
(see, e.g., \cite%
{Savage,SavageKheruntsyanSpatial,Deuar-Drummond-PRL-2007,Midgley,Magnus-Karen-4WM,Perrin-theory,KK-Radius}%
), we restrict our study only to dissociation into bosonic atoms.
Development of similar techniques for fermions is underway \cite%
{Corney-fermionic,Magnus-Karen-Joel-First-Principles}, but they are so far
limited to treating homogeneous systems and therefore are not adopted yet to
the present problem of dissociation of spatially inhomogeneous molecular
condensates.

To understand the role molecular depletion by itself, we first treat
the dissociation dynamics governed by the Hamiltonian (\ref{hamiltonian})
without the $s$-wave scattering interactions. Since most of our previous
numerical and analytic examples were given for the case of distinguishable
atoms (which has its fermionic counterpart), we examine the role of
molecular depletion in the same examples.

In the second part of this section we analyze the role of collisional
interactions between the atoms and molecules and restrict our study to
the case of indistinguishable bosonic atoms. More specifically, we
treat molecule-molecule, molecule-atom, and atom-atom $s$-wave scattering
interactions described by the respective $s$-wave scattering lengths $a_{00}$%
, $a_{01}$, and $a_{11}$. Our treatment automatically incorporates
molecular depletion since the molecular field in all cases is treated
quantum mechanically, without invoking the mean-field approximation. The
restriction to the case of indistinguishable atoms is motivated by the need
to keep the parameter space manageable, while still giving us an overall
quantitative understanding of the role of these collisional processes. For
comparison, in the case of dissociation into distinguishable atoms, a
generic treatment would have to incorporate six different types of intra-
and interspecies scattering processes described by six (generally
different, and yet unknown for most of the species) scattering lengths $%
a_{ij}$ ($i,j=0,1,2$), which is a challenging task and is beyond the scope
of the present article.

\subsection{Role of molecular depletion}

\label{sec:The-role-of-depletion} To model the quantum dynamics of
dissociation beyond the undepleted molecular-field approximation, we use a
first-principles phase-space method based on the positive-$P$
representation. In this method, each pair of the field operators $\hat{\Psi}%
_{i}(x,t)$ and $\hat{\Psi}_{i}^{\dagger }(x,t)$ ($i=0,1,2$) in the
Hamiltonian (\ref{hamiltonian}) in 1D is represented by two complex
stochastic fields $\Psi _{i}(x,t)$ and $\tilde{\Psi}_{i}(x,t)$ whose
dynamics is governed by a set of stochastic differential equations,%
\begin{eqnarray*}
\frac{\partial \Psi _{0}}{\partial t} &=&i\frac{\hbar }{2m_{0}}\frac{%
\partial ^{2}\Psi _{0}}{\partial x^{2}}-\chi \Psi _{1}\Psi _{2}, \\
\frac{\partial \Psi _{1}}{\partial t} &=&i\left[ \frac{\hbar }{2m_{1}}\frac{%
\partial ^{2}}{\partial x^{2}}-\Delta \right] \Psi _{1}+\chi \Psi _{0}\tilde{%
\Psi}_{2}+\sqrt{\chi \Psi _{0}}\xi _{1}, \\
\frac{\partial \Psi _{2}}{\partial t} &=&i\left[ \frac{\hbar }{2m_{2}}\frac{%
\partial ^{2}}{\partial x^{2}}-\Delta \right] \Psi _{2}+\chi \Psi _{0}\tilde{%
\Psi}_{1}+\sqrt{\chi \Psi _{0}}\xi _{1}^{\ast },
\end{eqnarray*}%
\begin{eqnarray}
\frac{\partial \tilde{\Psi}_{0}}{\partial t} &=&-i\frac{\hbar }{2m_{0}}\frac{%
\partial ^{2}\tilde{\Psi}_{0}}{\partial x^{2}}-\chi \tilde{\Psi}_{1}\tilde{%
\Psi}_{2},  \notag \\
\frac{\partial \tilde{\Psi}_{1}}{\partial t} &=&-i\left[ \frac{\hbar }{2m_{1}%
}\frac{\partial ^{2}}{\partial x^{2}}-\Delta \right] \tilde{\Psi}_{1}+\chi
\tilde{\Psi}_{0}\Psi _{2}+\sqrt{\chi \tilde{\Psi}_{0}}\xi _{2},  \notag \\
\frac{\partial \tilde{\Psi}_{2}}{\partial t} &=&-i\left[ \frac{\hbar }{2m_{1}%
}\frac{\partial ^{2}}{\partial x^{2}}-\Delta \right] \tilde{\Psi}_{2}+\chi
\tilde{\Psi}_{0}\Psi _{1}+\sqrt{\chi \tilde{\Psi}_{0}}\xi _{2}^{\ast }.
\notag \\
&&\;
\end{eqnarray}%
Here $\xi _{1}=( \zeta _{1}+i\zeta _{2} ) / \sqrt{2} $ and $\xi _{2}=(
\zeta _{3}+i\zeta _{4} ) / \sqrt{2}$ are the complex noise terms, with $%
\zeta _{j}(x,t)$ ($j=1,2,3,4$) being real independent Gaussian noises with
zero mean, $\left\langle \zeta _{j}(x,t)\right\rangle =0$, and the following
nonzero correlations: $\left\langle \zeta _{i}(x,t)\zeta _{j}(x^{\prime
},t^{\prime })\right\rangle =\delta _{ij}\delta (x-x^{\prime })\delta
(t-t^{\prime })$. The stochastic fields $\Psi _{i}(x,t)$ and $\tilde{\Psi}%
_{i}(x,t)$ are independent of each other [$\tilde{\Psi}_{i}(x,t)\neq \Psi
_{i}^{\ast }(x,t)$, $t>0$], except in the mean, $\langle \tilde{\Psi}%
_{i}(x,t)\rangle =\langle \Psi _{i}^{\ast }(x,t)\rangle $, where the
brackets refer to stochastic averages with respect to the positive-$P$
distribution function. In numerical realizations, this is represented by an
ensemble average over a large number of stochastic realizations
(trajectories). Observables described by quantum mechanical ensemble
averages over normally ordered operator products have an exact
correspondence with stochastic averages over the fields $\Psi (x,t)$ and $%
\tilde{\Psi}(x,t)$.

The initial condition for our simulations is a vacuum state for the atomic
fields and a coherent state for the molecular condensate, with $\tilde{\Psi}%
_{0}(x,0)=\Psi _{0}^{\ast }(x,0)$. In the numerical examples, we assume that
the molecular condensates initially have the same density profiles as those
used in our simulations within the undepleted molecular approximation (see
Fig. \ref{figDifferentWidths}). We note that in the undepleted case, these
density profiles were assumed to originate from the ground-state solution of
the Gross-Pitaevskii equation in a harmonic trap, which in the TF limit
gives inverted parabolas whose TF radii depend on the strength of
molecule-molecule interactions. In the present case, we treat the molecular
depletion but ignore the molecule-molecule $s$-wave scattering interactions;
accordingly, the ground state of a harmonic trap would result in a Gaussian
density profile rather than an inverted parabola. To make the present
treatment self-consistent, yet directly comparable with the numerical
examples analyzed in the undepleted case, we therefore assume that the same,
near-parabolic initial density profiles are created by tailoring the shape
of the trapping potential.

The results of the exact positive-$P$ simulations are shown in Figs.~\ref%
{total-atom-number}, \ref{peak-correlation}, and \ref{variance}; the
comparison with the results of the undepleted molecular approximation is
discussed in the respective parts of text in Sec.~\ref{sec:Numerics}. A
known drawback of the positive-$P$ method is that it suffers from
increasingly large sampling errors due to the boundary terms problem \cite%
{Gilchrist,Deuar-Drummond-2006} as the simulation time increases, eventually
leading to diverging results. In the present examples without the $s$-wave
scattering terms, the useful simulation times were limited to $t\simeq
6.5t_{0}$.

In order to go beyond the simulation durations achievable via the positive-$%
P $ method, we have also performed simulations using the truncated
Wigner-function approach. Unlike the positive-$P$ method, it is an
approximate approach as it involves neglecting or truncating third- and
higher-order derivative terms in the evolution equation for the Wigner
function. This is necessary in order to obtain an equation in the form of a
Fokker-Planck equation which can then be mapped onto a set of stochastic
differential equations. These equations formally render as deterministic
mean-field (or Gross-Pitaevskii-like) equations,%
\begin{eqnarray}
\frac{\partial \Psi _{0}}{\partial t} &=&i\frac{\hbar }{2m_{0}}\frac{%
\partial ^{2}\Psi _{0}}{\partial x^{2}}-\chi \Psi _{1}\Psi _{2},  \notag \\
\frac{\partial \Psi _{1}}{\partial t} &=&i\left[ \frac{\hbar }{2m_{1}}\frac{%
\partial ^{2}}{\partial x^{2}}-\Delta \right] \Psi _{1}+\chi \Psi _{0}\Psi
_{2}^{\ast },  \notag \\
\frac{\partial \Psi _{2}}{\partial t} &=&i\left[ \frac{\hbar }{2m_{2}}\frac{%
\partial ^{2}}{\partial x^{2}}-\Delta \right] \Psi _{2}+\chi \Psi _{0}\Psi
_{1}^{\ast };
\end{eqnarray}%
however, their stochastic nature and quantum fluctuations are included by
way of a noise contribution in the initial state for the molecular and
atomic fields. The addition of this initial vacuum noise ensures that the
initial states of $\Psi _{0}$ and $\Psi _{1,2}$ represent the Wigner
function of an initial coherent-state BEC for the molecules and an initial
vacuum state for the atoms, respectively. The corresponding stochastic
averages with the Wigner distribution function correspond to symmetrically
ordered operator products, so that the calculation of observables
represented by normally ordered operator products needs appropriate
symmetrization.

The results of our simulations using the Wigner function method are shown in
Figs.~\ref{total-atom-number} and \ref{peak-correlation} and are in
excellent agreement with the exact positive-$P$ results, thus reinforcing
our confidence in the adequacy of this method for treating the problem of
molecular dissociation.

\subsection{Role of collisional interactions}

We now turn to the treatment of $s$-wave scattering interactions in the case
of dissociation into indistinguishable bosonic atoms. This case is described
by the Hamiltonian (\ref{Hindistinguish}), together with the additional
quartic terms,
\begin{equation}
\widehat{H}_{int}=\sum\limits_{i,j=0,1}\frac{\hbar U_{ij}^{(1D)}}{2}\int dx\,%
\widehat{\Psi }_{i}^{\dagger }\widehat{\Psi }_{j}^{\dagger }\widehat{\Psi }%
_{j}\widehat{\Psi }_{i}.  \label{Hint}
\end{equation}%
Here $U_{00}^{(1D)}$, $U_{01}^{(1D)}=U_{10}^{(1D)}$, and $U_{11}^{(1D)}$
correspond to the effective 1D coupling constants describing, respectively,
molecule-molecule, molecule-atom, and atom-atom interactions that are
proportional to the 3D scattering lengths $a_{00}$, $a_{01}$, and $a_{11}$.
In the case of a harmonic transverse confinement realizing a cigar-shaped 1D
system, these constants are given by $U_{ii}^{(1D)}=2\omega _{\bot }a_{ii}$
and $U_{01}^{(1D)}=(3/\sqrt{2})\omega _{\bot }a_{01}$, where $\omega _{\bot
} $ is the transverse harmonic oscillator frequency \cite{PhysicalParameters2}.

The treatment of the $s$-wave scattering interactions using the positive-$P$
method is a challenging task because the problem of growing sampling errors
becomes more severe than before. The useful simulation time with realistic
physical parameters reduces to sub-milliseconds in our examples, which is
too short to give any new insights beyond the undepleted molecular
approximation (see also Ref. \cite{Savage}). As discussed in Ref. \cite%
{Midgley}, the most reliable method in this situation is the truncated
Wigner approach, which produces stochastic equations that remain stable for
much longer simulation durations than the positive-$P$ equations. The Wigner
equations in the present case are%
\begin{eqnarray}
\frac{\partial \Psi _{0}}{\partial t} &=&i\frac{\hbar }{2m_{0}}\frac{%
\partial ^{2}\Psi _{0}}{\partial x^{2}}-i\sum_{j=0,1}U_{0j}|\Psi
_{j}|^{2}\Psi _{0}-\frac{\chi }{2}\Psi _{1}^{2},  \label{WEq} \\
\frac{\partial \Psi _{1}}{\partial t} &=&i\left[ \frac{\hbar }{2m_{1}}\frac{%
\partial ^{2}}{\partial x^{2}}-\Delta -\sum_{j=0,1}U_{1j}|\Psi _{j}|^{2}%
\right] \Psi _{1}+\chi \Psi _{0}\Psi _{1}^{\ast },  \notag
\end{eqnarray}%
and the quantum fluctuations are introduced as previously via the noise
contributions in the initial state for the molecular and atomic fields.

\begin{figure}[tbph]
\includegraphics[height=5cm]{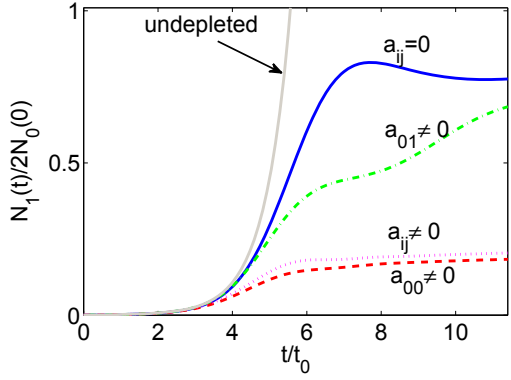} 
\caption{(Color online) Fractional total atom number as a function of time, $%
N_{1}(t)/2N_{0}(0)$, in dissociation into indistinguishable bosonic atoms.
The timescale $t_{0}=1/\protect\chi \protect\sqrt{\protect\rho _{0}}$ in
these examples is $t_{0}\simeq 0.035$ s \protect\cite{PhysicalParameters2}.
The two solid curves (light gray and blue) are reference examples,
corresponding to the undepleted molecular approximation (labeled) and the
case with molecular depletion but no $s$-wave scattering interactions ($%
a_{ij}=0$ nm). The remaining three curves correspond to simulations in which
we include: molecule-molecule interactions with $a_{00}=3$ nm (and $%
a_{01}=a_{11}=0$), dashed red curve; molecule-atom interactions with $%
a_{01}=3$ nm (and $a_{00}=a_{11}=0$), dash-dotted green curve; and all
three $s$-wave scattering interactions with $a_{00}=a_{01}=a_{11}=3$ nm,
dotted magenta curve. In the undepleted case, we assume a near-parabolic
initial density profile of the molecular BEC, with the peak 1D density $%
\protect\rho _{0}\simeq 1.8\times 10^{7}$ m$^{-1}$ and a radius $%
R_{TF}\simeq 53$ $\protect\mu $m in the TF limit (originating from a
transverse trapping potential frequency of $\protect\omega _{\bot }/2\protect%
\pi =54$ Hz and molecule-molecule 3D scattering length of $a_{00}=3$ nm). In
the cases with depletion but no $s$-wave scattering interactions, we assume a
Gaussian density profile $\protect\rho _{0}(x,0)=\protect\rho _{0}\exp
(-x^{2}/2S_{x}^{2})$ that has the same peak 1D density and an rms width of $%
S_{x}=35$ $\protect\mu $m which is chosen as to closely follow the
near-parabolic profile for the interacting case in the central part of the
cloud. The same near-parabolic initial density profile was used in the
examples with $a_{00}=3$ nm (dashed red and dotted magenta), and the same
Gaussian profile was used in the example with $a_{01}=3$ nm and $%
a_{00}=a_{11}=0$ nm (dash-dotted green).}
\label{fig:NatomsWigner}
\end{figure}

As an example relevant to practice, we treat $^{87}$Rb$_{2}$ molecules as in
the experiments of Ref. \cite{Rempe-diss} and use the 1D parameters
described in \cite{PhysicalParameters2}. In Fig.~\ref{fig:NatomsWigner} we
plot the total number of free dissociated atoms $N_{1}(t)$ relative to their
initial number within the molecular condensate $2N_{0}(0)$. The two solid
curves are reference examples, corresponding to the undepleted molecular
approximation (light gray) and the case when the molecular depletion is
included, but all $s$-wave scattering interactions are still absent (blue
curve). The dashed and the dash-dotted curves are, respectively, for the
cases that include only molecule-molecule and molecule-atom interactions
(the case with only atom-atom interactions was studied in \cite{Midgley}),
while the dotted curve corresponds to the inclusion of all three types of $s$%
-wave scattering terms. As we see, all these curves agree with the curve for
the undepleted molecular approximation for durations of $t/t_{0}\lesssim 3$.
We therefore conclude that the inclusion of $s$-wave scattering interactions
reduces the regime of validity of the undepleted molecular approximation to
durations corresponding approximately to $5\%$ conversion, which is about
twice lower than in the absence of these interactions.

In order to qualitatively understand the behavior of the different curves in
Fig. \ref{fig:NatomsWigner}, we note the following two dominant features: (1)
the two curves (dashed red and dotted magenta) that correspond to having
nonzero molecule-molecule interactions grow slower than the other curves,
and (2) all curves with $s$-wave scattering interactions \textquotedblleft bend\textquotedblright\ at around $%
t/t_{0}\sim 5$ and do not reach the maximum conversion efficiency seen in
the solid blue curve.

\begin{figure}[tbp]
\includegraphics[width=8.7cm]{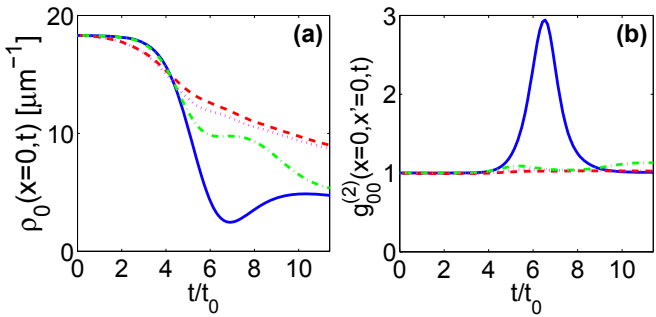}
\caption{(Color online) (a) Peak molecular density $\protect\rho _{0}(0,t)$
as a function of time. The different curves are as in Fig. \protect\ref%
{fig:NatomsWigner}. (b) Molecule-molecule local pair correlation in the
center of the molecular cloud $g_{00}^{(2)}(x=0,x^{\prime }=0,t)$ as a
function of time, for the same curves as in (a).}
\label{fig:molecular-density}
\end{figure}

The slower growth of the curves that include the effect of molecule-molecule
interactions is explained by the fact that the molecular condensate in these
examples experiences additional expansion due to the repulsive $s$-wave
scattering interactions. Such an expansion is accompanied by a faster
reduction of the molecular peak density [see Fig. \ref{fig:molecular-density}%
(a)] compared to the cases with no molecule-molecule interactions. Indeed,
as we see in Fig. \ref{fig:molecular-density}(a) the molecular peak density
for the dashed red and dotted magenta curves drops faster than for the other
two curves, yet the atom number growths for the same cases is the slowest in
Fig. \ref{fig:NatomsWigner}. Thus, the reduction in the molecular density
takes place not only because of the conversion to atoms, but also because of
the expansion of the molecular condensate. Our estimates show that at $%
t/t_{0}\simeq 3$ the simple expansion alone would reduce the molecular
density to $0.95$ of the original value, which is a larger effect than the
reduction due to conversion to atoms. In terms of an instantaneous effective
coupling $\chi \langle \hat{\Psi}_{0}(x,t) \rangle $ (interpreted at the
level of a time-dependent mean field), which is similar to $g(x)=\chi \sqrt{%
\rho _{0}(x,0)}$ used previously in the undepleted molecular approximation,
the additional reduction in the molecular density means that the
instantaneous atom-molecule coupling and therefore the rate of conversion
into atoms is reduced more than in the case with no expansion of the
molecular condensate. Accordingly, the dashed red and dotted magenta curves
in Fig. \ref{fig:NatomsWigner} have the slowest growth in the atom number.

The \textquotedblleft bending\textquotedblright\ of the curves at later times can be explained by the effect of
phase diffusion due to the $s$-wave scattering interactions. The phase
diffusion leads to dynamical dephasing of the phase-matching condition for
efficient conversion, thus suppressing an exponential amplification in the
atom number growth. The characteristic time scale for phase diffusion in the
initial stages of dissociation can be estimated via
\begin{equation}
t_{d}\simeq \frac{2\pi }{|U_{01}^{(1D)}-\frac{1}{2}U_{00}^{(1D)}|\rho _{0}},
\end{equation}%
which gives the following results for the three cases shown in Fig. \ref%
{fig:NatomsWigner}: $t_{d}/t_{0}\simeq 4.6$\ for the case with $a_{01}\neq 0$%
, $t_{d}/t_{0}\simeq 9.8$\ for the case with $a_{00}\neq 0$, and $%
t_{d}/t_{0}\simeq 8.7$\ for the case with all $s$-wave scattering
interactions present. Here we have ignored the contribution coming from the
atomic mean field itself (as the atomic density is initially negligibly
small compared to the molecular peak density) and ignored the fact that the
mean-field phase shifts are spatially dependent and dynamically changing.
With these remarks in mind, we find that our order-of-magnitude estimates of
the phase diffusion time are consistent with the numerical results seen in
Fig. \ref{fig:NatomsWigner}. Moreover, a model simulation of a uniform system
with the same $s$-save scattering interaction terms as in the relevant
three cases of Fig. \ref{fig:NatomsWigner} reproduce the trend that follows
from these simple estimates and the order of the curves that experience
progressively slower phase diffusion.

In addition to monitoring the dynamics of the molecular peak density, we
have analyzed the second-order correlation function $g_{00}^{(2)}(x,x^{%
\prime },t)$ for the molecular field in position space in order to
understand the deviation of the molecular field from the initial coherent
state. In Fig. \ref{fig:molecular-density}(b) we show the molecule-molecule
local pair correlation in the center of the cloud $g_{00}^{(2)}(0,0,t)$ as a
function of time for the same examples as in Fig. \ref{fig:molecular-density}%
(a). We see that the strongest deviation from the coherent state value of $%
g_{00}^{(2)}(0,0,t)=1$ occurs for the case with the strongest depletion,
corresponding to the absence of $s$-wave scattering interactions (solid blue
curve). In this case, the minimum in the peak molecular density---occurring
approximately at $t/t_{0}\sim 6.5$---is the smallest and is closer to zero,
implying that the quantum fluctuations are no longer negligible compared to
the mean-field part. As a signature of this, we see a respective peak in the
pair correlation function, $g_{00}^{(2)}(0,0,t)\simeq 3$, around the same
time. In contrast to this, in cases with $s$-wave scattering interactions,
the molecular depletion is weaker, the peak density is still high enough,
and therefore the molecular field remains closer to the initial coherent
state in terms of the second-order coherence, with $g_{00}^{(2)}(0,0,t)%
\simeq 1$.

\begin{figure}[tbp]
\includegraphics[height=5.5cm]{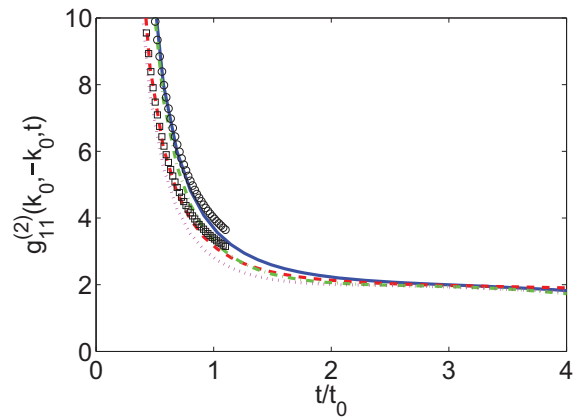}
\caption{(Color online) Back-to-back pair correlation for dissociated atoms,
$g_{11}^{(2)}(k_{0},-k_{0},t)$, as a function of time. The different curves
(solid blue, dash-dotted green, dashed red, and dotted magenta) are as in
Fig. \protect\ref{fig:NatomsWigner}. The black squares and circles are the
results from short-time analytic solutions given by Eqs. (\protect\ref%
{g_11IndistBB1DTF}) and (\protect\ref{g_11IndistBBGaussian}), respectively.}
\label{fig:corr}
\end{figure}

Finally, in Fig. \ref{fig:corr} we plot the back-to-back pair correlation
function for the dissociated atoms in momentum space and monitor it as a
function of time. The different curves (solid, dashed, dash-dotted, and
dotted) are for the same numerical examples as in Fig. \ref{fig:NatomsWigner}%
, showing that the $s$-wave scattering interactions---while certainly
affecting the quantitative details---have a relatively smaller effect on
atom-atom correlation functions at least for dissociation durations up to $%
t/t_{0}\simeq 4$. Moreover, the numerical results for $t/t_{0}\lesssim 1$
are in surprisingly good agreement with the short-time analytic results
(shown as squares for the case of the initial TF density profile of the
molecular BEC and as circles for the initial Gaussian profiles), which in
the strict sense were meant to be applicable only for $t/t_{0}\ll 1$. The
analytic results for the present case of indistinguishable bosons follow
from Eq. (\ref{g_11Indist}), which for $k^{\prime }=-k$ reads as $%
g_{11}^{(2)}(k,-k,t)=2+|m(k,-k,t)|^{2}/[n(k,t)]^{2}$. Using the asymptotic
expressions in 1D given by Eqs. (\ref{m1D}) and (\ref{n1D}) for the case of
the TF parabola, we obtain%
\begin{equation}
g_{11}^{(2)}(k,-k,t)\simeq 2+\frac{9\pi ^{2}}{64\chi ^{2}t^{2}\rho _{0}}.
\label{g_11IndistBB1DTF}
\end{equation}%
Similarly, using Eqs. (\ref{m1D_Gaussian}) and (\ref{n1D_Gaussian}) for the
Gaussian density profile, we obtain
\begin{equation}
g_{11}^{(2)}(k,-k,t)\simeq 2+\frac{2}{\chi ^{2}t^{2}\rho _{0}}.
\label{g_11IndistBBGaussian}
\end{equation}%
Both these expressions have inverse square dependence on time $t$ and are
shown in Fig. \ref{fig:corr} by the curves corresponding to squares and
circles.

\section{Summary}

\label{sec:Summary}

We have analyzed the quantum dynamics of dissociation of harmonically
trapped Bose-Einstein condensates of molecular dimers in one, two, and three
spatial dimensions. More specifically, we have initially examined how the
spatial inhomogeneity of the molecular condensate affects the conversion
dynamics and the atom-atom correlations in momentum space in the short-time
limit. Both fermionic and bosonic statistics of the constituent atom pairs
were considered. Using the undepleted molecular-field approximation, we have
obtained explicit analytic results for the short-time asymptotic behavior of
the pair correlation functions and the relative atom number squeezing.

For a TF parabolic density profile of the molecular BEC, the correlation
functions can be expressed in terms of Bessel functions and are determined
by the momentum distribution of the source molecular BEC. The precise
relationship is different in one, two, and three dimensions. We have
compared the results corresponding to the TF density profile with those
corresponding to a simple Gaussian and found that the correlation widths
relative to the source width are smaller in the TF case. We have also shown
that the relative atom number squeezing in a given atom counting volume is
determined merely by the characteristic size of the molecular BEC and by a
shape-dependent geometric factor and that the squeezing improves as the
size of the BEC is increased. The strength of atom-atom correlations for
equal but opposite momenta and the relative number squeezing are the
strongest in 1D systems, where the mode-mixing is a relatively weaker effect
than in higher dimensions.

The analytic and numerical results in the undepleted molecular-field
approximation have been compared with exact numerical simulations using the
positive-$P$ representation in 1D, as well as with simulations using the
truncated Wigner approach, both of which are suitable for treating the
molecular-field dynamics and its depletion. The comparison shows that in the
absence of any $s$-wave scattering interactions the undepleted molecular
approximation is valid for dissociation durations corresponding to $\sim
10\% $ conversion of molecules into free atoms. We have also examined the
role of the $s$-wave scattering interactions within the truncated Wigner
approach and found that the phase diffusion due to these interactions can
significantly modify the conversion dynamics in the long time limit.
Nevertheless, for dissociation durations corresponding to $\sim 5\%$
conversion, one can still neglect these interactions and use the undepleted
molecular approximation. Furthermore, our results indicate that a possible
route to extending the validity of the undepleted molecular approximation
and minimizing the role of collisional interactions is to operate at larger
absolute values of both the atom-molecule coupling $\chi $ and the
dissociation detuning $|\Delta |$.

Even though $5\%-10\%$ conversion efficiencies seem small, nevertheless they
can produce mesoscopic ensembles of pair-correlated atoms with interesting
quantum statistics and nontrivial many-body correlations if one starts with
large-enough molecular condensates, such as containing $10^{4}-10^{5}$
molecules in typical 1D configurations. Accordingly, our simple analytic
results in the undepleted molecular approximation provide a useful tool for
obtaining qualitative insights and better than \textquotedblleft order-of-magnitude\textquotedblright\
estimates for realistic inhomogeneous systems. For a specific system at
hand, these results can be further refined using the positive-$P$ and
truncated Wigner methods as demonstrated in the present study.

\begin{acknowledgments}
The authors thank C. M. Savage for useful discussions and acknowledge
support by the Australian Research Council through the ARC Centre of
Excellence scheme. MÖ acknowledges support by IPRS/UQILAS and the Solander
Program at UQ.
\end{acknowledgments}

\appendix

\section{Correlation integrals in 2D}

\label{2D-integrals}

We first evaluate the integral in the anomalous density $|m_{12}(\mathbf{k},%
\mathbf{k}^{\prime },t)|$ [Eq. (\ref{m12})], in which we take $\mathbf{k}%
^{\prime }=-\mathbf{k+e}_{i}\Delta k_{i}$, so that the integral takes the
form
\begin{gather}
|m_{12}(k_{x},k_{x}^{\prime },t)|=\frac{t\chi \sqrt{\rho _{0}}}{(2\pi )^{2}}%
\int\limits_{\Lambda }dx\,dy  \notag \\
\times e^{i(k_{x}+k_{x}^{\prime })x}\left( 1-\frac{x^{2}}{R_{\mathrm{TF}%
,x}^{2}}-\frac{y^{2}}{R_{\mathrm{TF},y}^{2}}\right) ^{1/2},  \label{a1}
\end{gather}%
where the integration domain $\Lambda $ is given by $x^2/R_{\mathrm{TF}%
,x}^{2}+y^2/R_{\mathrm{TF},y}^{2}<1$. For definiteness, we have chosen the
direction $i$ to be the $x$ axis, without the loss of generality.
Introducing scaled variables, $x^{\prime }=x/R_{\mathrm{TF},x}$ and $%
y^{\prime }=y/R_{\mathrm{TF},y}$, and transforming to polar coordinates $%
x^{\prime }=r\cos \theta $ and $y^{\prime }=r\sin \theta $, the integral can
be rewritten as
\begin{gather}
|m_{12}(k_{x},k_{x}^{\prime },t)|=\frac{t\chi \sqrt{\rho _{0}}R_{\mathrm{TF}%
,x}R_{\mathrm{TF},y}}{(2\pi )^{2}}  \notag \\
\times \int\limits_{0}^{1}dr\,r\,(1-r^{2})^{1/2}\int\limits_{0}^{2\pi
}d\theta \,e^{i(k_{x}+k_{x}^{\prime })R_{\mathrm{TF},x}r\cos \theta }.
\label{a2}
\end{gather}

Using the integral representation of the zeroth-order Bessel function \cite%
{Bateman}
\begin{equation}
J_{0}(z)=\frac{1}{\pi }\int\limits_{0}^{\pi }d\theta \,e^{iz\cos \theta },
\label{a3}
\end{equation}%
the integral in Eq. (\ref{a2}) can be brought into the following form:%
\begin{gather}
|m_{12}(k_{x},k_{x}^{\prime },t)|=\frac{t\chi \sqrt{\rho _{0}}R_{\mathrm{TF}%
,x}R_{\mathrm{TF},y}}{2\pi }  \notag \\
\times \int\limits_{0}^{1}dr\,r\,(1-r^{2})^{1/2}J_{0}\left(
(k_{x}+k_{x}^{\prime })R_{\mathrm{TF},x}\ r\right)  \label{a4}
\end{gather}

By making a variable change $r=\sin \phi $ and using the integral
representation of the Bessel function of general order \cite{Bateman},
\begin{eqnarray}
&&J_{\rho +\mu +1}(q)\,2^{\rho }\,\Gamma (\rho +1)\,q^{-\rho -1}  \notag \\
&=&\int\limits_{0}^{\pi /2}d\phi J_{\mu }(q\sin \phi )(\sin \phi )^{\mu
+1}(\cos \phi )^{2\rho +1},  \label{a5}
\end{eqnarray}%
with $\mathrm{Re}\,\rho >-1$ and $\mathrm{Re}\,\mu >-1$, we finally obtain
\begin{gather}
|m_{12}(k_{x},k_{x}^{\prime },t)|\simeq \frac{t\chi \sqrt{\rho _{0}}R_{%
\mathrm{TF},x}R_{\mathrm{TF},y}}{2\sqrt{2\pi }}\;  \notag \\
\times \frac{J_{3/2}\left( (k_{x}+k_{x}^{\prime })R_{\mathrm{TF},x}\right) }{%
\left[ (k_{x}+k_{x}^{\prime })R_{\mathrm{TF},x}\right] ^{3/2}}.  \label{a6}
\end{gather}

If the displacement direction $i$ was along $y$, we would obtain the same
results except $k_{x}\rightarrow k_{y}$ by defining the polar coordinates
according to $x^{\prime }=r\sin \theta ^{\prime }$ and $y^{\prime }=r\cos
\theta ^{\prime }$. Thus, by replacing in $|m_{12}(k_{x},k_{x}^{\prime },t)|$
the $k_{x}$ and $k_{x}^{\prime }$ components with $k_{i}$ and $k_{i}^{\prime
}$ ($i=x,y$) , we arrive at the general result of Eq. (\ref{m12-2D}).

To evaluate the integrals in the normal density $n_{j}(\mathbf{k},\mathbf{k}%
^{\prime },t)$ [Eq. (\ref{njj})], in which we take $\mathbf{k}^{\prime }=%
\mathbf{k+e}_{i}\Delta k_{i}$, and therefore
\begin{gather}
n_{j}(k_{x},k_{x}^{\prime },t)=\frac{t^{2}\chi ^{2}\rho _{0}}{(2\pi )^{2}}%
\int\limits_{\Lambda }dx\,dy  \notag \\
\times e^{i(k_{x}-k_{x}^{\prime })x}\left( 1-\frac{x^{2}}{R_{\mathrm{TF}%
,x}^{2}}-\frac{y^{2}}{R_{\mathrm{TF},y}^{2}}\right) ,  \label{a7}
\end{gather}%
we follow the same steps as in evaluating $|m_{12}(k_{x},k_{x}^{\prime },t)|$%
. In polar coordinates, the preceding integral can be brought into the following
form:%
\begin{gather}
n_{j}(k_{x},k_{x}^{\prime },t)=\frac{t^{2}\chi ^{2}\rho _{0}}{(2\pi )^{2}}
\notag \\
\times \int\limits_{0}^{1}dr\,r\,(1-r^{2})\int\limits_{0}^{2\pi }d\theta
\,e^{i(k_{x}-k_{x}^{\prime })R_{\mathrm{TF},x}r\cos \theta }  \notag \\
=\frac{t^{2}\chi ^{2}\rho _{0}R_{\mathrm{TF},x}R_{\mathrm{TF},y}}{2\pi }%
\int\limits_{0}^{1}dr\,r\,(1-r^{2})J_{0}\left( (k_{x}-k_{x}^{\prime })R_{%
\mathrm{TF},x}\ r\right) .  \label{a8}
\end{gather}

Introducing $r=\sin \phi $ and using Eq. (\ref{a5}), the integration with
respect to $\phi $ gives
\begin{equation}
n_{j}(k_{x},k_{x}^{\prime },t)\simeq \frac{t^{2}\chi ^{2}\rho _{0}R_{\mathrm{%
TF},x}R_{\mathrm{TF},y}}{\pi }\frac{J_{2}\left( (k_{x}-k_{x}^{\prime })R_{%
\mathrm{TF},x}\right) }{\left[ (k_{x}-k_{x}^{\prime })R_{\mathrm{TF},x}%
\right] ^{2}}.
\end{equation}%
The same result, except $k_{x}\rightarrow k_{y}$, can be obtained if the
displacement is along $y$, and we arrive at the general result of Eq. (\ref%
{njj-2D}).

Finally, the calculation of the integral in
\begin{equation}
n_{0}(k_{i})=\left\vert \int d^{2}\mathbf{x}\sqrt{\rho_{0} (\mathbf{x})}\exp
(-ik_{i}x_{i})/(2\pi )\right\vert ^{2}
\end{equation}%
follows the same pattern as in $|m_{12}(k_{i},k_{i}^{\prime },t)|$ except
that the relevant term is squared, and one obtains Eq. (\ref{2Dmolmomdist}).

\section{Correlation integrals in 3D}

\label{3D-integrals}

We first evaluate the 3D integral in the anomalous density $|m_{12}(\mathbf{k%
},\mathbf{k}^{\prime },t)|$ [Eq. (\ref{m12})], in which we take $\mathbf{k}%
^{\prime }=-\mathbf{k+e}_{i}\Delta k_{i}$. For definiteness, we choose the
direction $i$ to be along $z$, so that the integral takes the form
\begin{gather}
|m_{12}(k_{z},k_{z}^{\prime },t)|=\frac{t\chi \sqrt{\rho _{0}}}{(2\pi )^{3}}%
\int\limits_{\Lambda }dx\,dy\,\,dz\,  \notag \\
\times e^{i(k_{z}+k_{z}^{\prime })z}\left( 1-\frac{x^{2}}{R_{\mathrm{TF}%
,x}^{2}}-\frac{y^{2}}{R_{\mathrm{TF},y}^{2}}-\frac{z^{2}}{R_{\mathrm{TF}%
,z}^{2}}\right) ^{1/2},
\end{gather}%
where the integration domain $\Lambda $ is defined by $x^{2}/R_{\mathrm{TF}%
,x}^{2}+y^{2}/R_{\mathrm{TF},y}^{2}+z^{2}/R_{\mathrm{TF},z}^{2}<1$.
Introducing scaled variables, $x^{\prime }=x/R_{\mathrm{TF},x}$, $y^{\prime
}=y/R_{\mathrm{TF},y}$, and $z^{\prime }=z/R_{\mathrm{TF},z}$ and
transforming to spherical coordinates $x^{\prime }=r\sin \theta \cos \varphi
$, $y^{\prime }=r\sin \theta \sin \varphi $, and $z^{\prime }=r\cos \theta $%
, the integral can be rewritten as
\begin{gather}
|m_{12}(k_{z},k_{z}^{\prime },t)|=\frac{t\chi \sqrt{\rho _{0}}R_{\mathrm{TF}%
,x}R_{\mathrm{TF},y}R_{\mathrm{TF},z}}{(2\pi )^{3}}  \notag \\
\times \int\limits_{0}^{2\pi }d\varphi
\int\limits_{0}^{1}dr\,r^{2}\,(1-r^{2})^{1/2}\int\limits_{0}^{\pi }d\theta
\sin \theta \,e^{i(k_{z}+k_{z}^{\prime })R_{\mathrm{TF},z}r\cos \theta }.
\end{gather}

The integral with respect to $\varphi $ is trivial, while the integral with
respect to $\theta $ is taken using \cite{Bateman}%
\begin{equation}
J_{\nu }(q)\Gamma (\nu +1/2)=\pi ^{-1/2}(q/2)^{\nu }\int\limits_{0}^{\pi
}d\theta (\sin \theta )^{2\nu }e^{iq\cos \theta },  \label{Bessel-another}
\end{equation}%
where $\mathrm{Re}\,\nu >-1/2$. This gives%
\begin{gather}
|m_{12}(k_{z},k_{z}^{\prime },t)|=\frac{t\chi \sqrt{\rho _{0}}R_{\mathrm{TF}%
,x}R_{\mathrm{TF},y}R_{\mathrm{TF},z}}{(2\pi )^{3/2}[(k_{z}+k_{z}^{\prime
})R_{\mathrm{TF},z}]^{1/2}}  \notag \\
\times \int\limits_{0}^{1}dr\,r^{3/2}\,(1-r^{2})^{1/2}J_{1/2}\left(
(k_{z}+k_{z}^{\prime })R_{\mathrm{TF},z}\ r\right) ,
\end{gather}%
which in turn takes the form of Eq. (\ref{a5}) if we introduce $r=\sin \phi $%
. Accordingly, we obtain%
\begin{gather}
|m_{12}(k_{z},k_{z}^{\prime },t)|=\frac{t\chi \sqrt{\rho _{0}}R_{\mathrm{TF}%
,x}R_{\mathrm{TF},y}R_{\mathrm{TF},z}}{4\pi }  \notag \\
\times \frac{J_{2}\left( (k_{z}+k_{z}^{\prime })R_{\mathrm{TF},z}\right) }{%
\left[ (k_{z}+k_{z}^{\prime })R_{\mathrm{TF},z}\right] ^{2}}.
\end{gather}

For displacements along $i=x$ or $i=y$ directions, one can obtain the same
result except that $k_{z}$ is replaced by $k_{i}$ by an appropriate rotation
of the spherical coordinate system, so that the final result for $%
|m_{12}(k_{i},k_{i}^{\prime },t)|$ takes the form of Eq. (\ref{m12-3D}).

To evaluate the integral in the normal density $n_{j}(\mathbf{k},\mathbf{k}%
^{\prime },t)$ [Eq. (\ref{njj})], in which $\mathbf{k}^{\prime }=\mathbf{k+e}%
_{i}\Delta k_{i}$, we follow the same steps as for evaluating $m_{12}(%
\mathbf{k},\mathbf{k}^{\prime },t)|$. In spherical coordinates, the integral
takes the following form:%
\begin{gather}
n_{j}(k_{z},k_{z}^{\prime },t)=\frac{t^{2}\chi ^{2}\rho _{0}R_{\mathrm{TF}%
,x}R_{\mathrm{TF},y}R_{\mathrm{TF},z}}{(2\pi )^{3}}  \notag \\
\times \int\limits_{0}^{2\pi }d\varphi
\int\limits_{0}^{1}dr\,r^{2}\,(1-r^{2})\int\limits_{0}^{\pi }d\theta \sin
\theta \,e^{i(k_{z}-k_{z}^{\prime })R_{\mathrm{TF},z}r\cos \theta }  \notag
\\
=\frac{t^{2}\chi ^{2}\rho _{0}R_{\mathrm{TF},x}R_{\mathrm{TF},y}R_{\mathrm{TF%
},z}}{(2\pi )^{3/2}[(k_{z}-k_{z}^{\prime })R_{\mathrm{TF},z}]^{1/2}}  \notag
\\
\times \int\limits_{0}^{1}dr\,r^{3/2}\,(1-r^{2})J_{1/2}\left(
(k_{z}-k_{z}^{\prime })R_{\mathrm{TF},z}\ r\right) ,
\end{gather}%
where we have used Eq. (\ref{Bessel-another}). Introducing $r=\sin \phi $
and using Eq. (\ref{a5}), we obtain%
\begin{gather}
n_{j}(k_{z},k_{z}^{\prime },t)=\frac{t^{2}\chi ^{2}\rho _{0}R_{\mathrm{TF}%
,x}R_{\mathrm{TF},y}R_{\mathrm{TF},z}}{\pi \sqrt{2\pi }}  \notag \\
\times \frac{J_{5/2}\left( (k_{z}-k_{z}^{\prime })R_{\mathrm{TF},z}\right) }{%
\left[ (k_{z}-k_{z}^{\prime })R_{\mathrm{TF},z}\right] ^{5/2}}.
\end{gather}%
Generalizing this to the arbitrary displacement direction $i$ leads to the
final result of Eq. (\ref{njj-3D}).

Finally, the calculation of the integral in
\begin{equation}
n_{0}(k_{i})=\left\vert \int d^{3}\mathbf{x}\sqrt{\rho_{0} (\mathbf{x})}\exp
(-ik_{i}x_{i})/(2\pi )^{3/2}\right\vert ^{2}
\end{equation}%
follows the same pattern as the evaluation of $|m_{12}(k_{i},k_{i}^{\prime
},t)|$, and we obtain Eq. (\ref{3Dmolmomdist}).


\begin{thebibliography}{99}
\bibitem{Comment2} In realistic experimental systems that are designed to
detect atom-atom correlations in position space (rather than in momentum
space) after time-of-flight expansion, further considerations arise which
are due to the initial position uncertainty of the molecules and the
detector resolution, as discussed in Ref. \cite{SavageKheruntsyanSpatial}.
\bibitem{PhysicalParameters} As a model system for physical parameters in
the case of fermionic atoms, we consider a 1D BEC of bosonic $^{40}$K$_{2}$
dimers ($m_{0}=2m=1.33\times 10^{-25}$ kg) having the peak 3D density $\rho
_{0}^{(3D)}=10^{20}$ m$^{-3}$ in the trap centre. Choosing the transverse
trap frequency $\omega _{\bot }/2\pi =1$ kHz gives the peak 1D density $\rho
_{0}\equiv \rho _{0}^{(1D)}=4\pi l_{\bot }^{2}\rho _{0}^{(3D)}=4\times
10^{7} $ m$^{-1}$, where $l_{\bot }=\sqrt{\hbar /m_{0}\omega _{\bot }}\simeq
0.355$ $\mu $m is the transverse harmonic oscillator length. Choosing next
the molecule-molecule $s$-wave scattering length (in 3D) to be $a_{00}=3$ nm
as an example, one can calculate the 1D coupling strength for the
molecule-molecule interaction, $U_{00}^{(1D)}\simeq 2\omega _{\bot }a$. In
the regime of week interactions [$\gamma \ll 1$, where $\gamma
=m_{0}U_{00}^{(1D)}/(\hbar ^{2}\rho _{0}^{(1D)})$], the Thomas-Fermi radius
of the molecular BEC in the longitudinal direction will be given by $R_{TF}=%
\sqrt{2\mu /m_{0}\omega _{x}^{2}}$, where $\mu =\hbar U_{00}^{(1D)}\rho
_{0}^{(1D)}$ is the chemical potential and $\omega _{x}$ is the longitudinal
trap frequency. In the case of distinguishable bosonic atoms, we assume the
same parameters -- as a hypothetical example -- to make the comparison with
the fermionic case directly comparable and dependent only on the quantum
statistics.
\bibitem{PhysicalParameters2} For the BEC of $^{87}$Rb$_{2}$ dimers (with $%
m_{1}=1.44\times 10^{-25}$ kg and $m_{0}=2m_{1}$) we consider a 3D density $%
\rho _{0}^{(3D)}=1.3\times 10^{18}$ m$^{-3}$ in the trap centre. Choosing
the transverse trap frequency $\omega _{\bot }/2\pi =54$ Hz gives the peak
1D density $\rho _{0}\equiv \rho _{0}^{(1D)}=4\pi l_{\bot }^{2}\rho
_{0}^{(3D)}\simeq 1.8\times 10^{7}$ m$^{-1}$, where $l_{\bot }=\sqrt{\hbar
/m_{0}\omega _{\bot }}\simeq 1.0$ $\mu $m is the transverse harmonic
oscillator length for the molecules. With all 3D $s$-wave scattering lengths
set to $a_{ij}=3$ nm, the 1D coupling strengths are $U_{ii}^{(1D)}=2\omega
_{\bot }a_{ii}\simeq 2.04\times 10^{-6}$ ms$^{-1}$ and $U_{01}=3/\sqrt{2}\
\omega _{\bot }a_{01}\simeq 2.16\times 10^{-6}$ ms$^{-1}$. We use a bare
detuning $\Delta =-259$ s$^{-1}$ and a 1D atom-molecule coupling $\chi
=6.7\times 10^{-3}$ m$^{1/2}$s$^{-1}$, resulting in the characteristic
timescale $t_{0}=1/\chi \sqrt{\rho _{0}}\simeq 0.035$ s. For the initial
density profiles in the examples of Fig. \ref{fig:NatomsWigner} with $%
a_{00}\neq 0$, we use a numerically determined (near-parabolic) ground state
solution of the GP equation for the molecular BEC in a longitudinal harmonic
trap with frequency $\omega _{x}/2\pi =1/2$ Hz ($\omega _{x}\ll \omega
_{\bot }$) and the same peak density $\rho _{0}$ as above. In the
Thomas-Fermi limit, the respective TF radius is given by $R_{TF}=\sqrt{%
4\hbar \omega _{\perp }a_{00}\rho _{0}/m_{0}\omega _{x}^{2}}\simeq 53\ \mu $%
m, and the corresponding initial total number of molecules is $%
N_{0}(0)=1.28\times 10^{3}$. For cases with a Gaussian initial density
profile, we use $\rho _{0}(x)=\rho _{0}\exp \left( -x^{2}/2S_{x}^{2}\right) $
with $S_{x}\simeq 35$ $\mu $m.
\end{thebibliography}
\end{document}